\documentclass[sigconf]{acmart}
\usepackage{amsmath}               
  {
      \theoremstyle{plain}
      \newtheorem{assumption}{Assumption}
  }
\usepackage{enumitem}
\AtBeginDocument{%
  \providecommand\BibTeX{{%
    \normalfont B\kern-0.5em{\scshape i\kern-0.25em b}\kern-0.8em\TeX}}}

\copyrightyear{2023} 
\acmYear{2023} 
\setcopyright{acmlicensed}\acmConference[KDD '23]{Proceedings of the 29th ACM SIGKDD Conference on Knowledge Discovery and Data Mining}{August 6--10, 2023}{Long Beach, CA, USA}
\acmBooktitle{Proceedings of the 29th ACM SIGKDD Conference on Knowledge Discovery and Data Mining (KDD '23), August 6--10, 2023, Long Beach, CA, USA}
\acmPrice{15.00}
\acmDOI{10.1145/3580305.3599462}
\acmISBN{979-8-4007-0103-0/23/08}

\usepackage{tikz}
\usepackage{subcaption}
\usepackage{pgfplots}
\usetikzlibrary{arrows,positioning,automata,calc,shapes}
\pgfplotsset{compat=newest, scaled z ticks=false} 
\pgfplotsset{plot coordinates/math parser=false}
\newlength\figureheight 
\newlength\figurewidth
\usepackage{caption}
\usepackage{filecontents}
\usepackage{url}
\usepackage{amsmath}
\usepackage{graphicx}
\usepackage{hyperref}

\usepackage{amsfonts,amssymb}
\usepackage{color, soul}
\usepackage{mathrsfs}
\usepackage{cleveref}
\usepackage{multirow}
\usepackage{wrapfig}
\usepackage{amsthm}
\usepackage{algorithm}
\usepackage{algorithmic}
\usepackage{bm}

\newcommand{\longeq}{\scalebox{2}[1]{=}}

\begin{document}

\title{Path-Specific Counterfactual Fairness for Recommender Systems}

\author{Yaochen Zhu}
\affiliation{%
\institution{University of Virginia}
\city{}
\state{}
\country{}
}
\email{uqp4qh@virginia.edu}

\author{Jing Ma}
\affiliation{%
\institution{University of Virginia}
\city{}
\state{}
\country{}
}
\email{jm3mr@virginia.edu}

\author{Liang Wu}
\affiliation{%
\institution{LinkedIn Inc.}
\city{}
\state{}
\country{}
}
\email{liawu@linkedin.com}

\author{Qi Guo}
\affiliation{%
\institution{LinkedIn Inc.}
\city{}
\state{}
\country{}
}
\email{qguo@linkedin.com}

\author{Liangjie Hong}
\affiliation{%
\institution{LinkedIn Inc.}
\city{}
\state{}
\country{}
}
\email{liahong@linkedin.com}

\author{Jundong Li}
\affiliation{%
\institution{University of Virginia}
\city{}
\state{}
\country{}
}
\email{jundong@virginia.edu}

\begin{abstract}
Recommender systems (RSs) have become an indispensable part of online platforms. With the growing concerns of algorithmic fairness, RSs are not only expected to deliver high-quality personalized content, but are also demanded not to discriminate against users based on their demographic information. However, existing RSs could capture undesirable correlations between sensitive features and observed user behaviors, leading to biased recommendations. Most fair RSs tackle this problem by completely blocking the influences of sensitive features on recommendations. But since sensitive features may also affect user interests in a fair manner (e.g., race on culture-based preferences), indiscriminately eliminating all the influences of sensitive features inevitably degenerate the recommendations quality and necessary diversities. To address this challenge, we propose a path-specific fair RS (PSF-RS) for recommendations. Specifically, we summarize all fair and unfair correlations between sensitive features and observed ratings into two latent proxy mediators, where the concept of path-specific bias (PS-Bias) is defined based on path-specific counterfactual inference. Inspired by Pearl's minimal change principle, we address the PS-Bias by minimally transforming the biased factual world into a hypothetically fair world, where a fair RS model can be learned accordingly by solving a constrained optimization problem. For the technical part, we propose a feasible implementation of PSF-RS, i.e., PSF-VAE, with weakly-supervised variational inference, which robustly infers the latent mediators such that unfairness can be mitigated while necessary recommendation diversities can be maximally preserved simultaneously. Experiments conducted on semi-simulated and real-world datasets demonstrate the effectiveness of PSF-RS.
\end{abstract}


%

\maketitle

\vspace{-3mm}

\section{Introduction}

As content grows exponentially on the web, recommender systems (RSs) are becoming increasingly critical in modern online service platforms \cite{zhang2019deep}. RSs capture user interests based on their historical behaviors \cite{he2017neural,ren2023disentangled}, profiles \cite{geng2015learning,zhu2022variational}, and the content of items they have interacted with \cite{yi2021cross,zhu2022mutually}, aiming to automatically deliver new items tailored to users' personalized interests. Nevertheless, the observed user behaviors may be unfairly correlated with certain sensitive user features, such as gender, race, and age, which can be unintentionally captured by the RSs and perpetuate into future recommendations \cite{li2021user}. Consequently, users may find the recommended items offensive, especially when people's concerns for discrimination have grown substantially over time \cite{mehrabi2021survey,obermeyer2019dissecting,ge2022survey,dong2023fairness}. 

In recent years, considerable efforts have been devoted to promoting fairness of RSs from both academia and industry  \cite{wang2022survey}. From the industry's perspective, several platforms are beginning to provide interfaces to encourage users to report potentially unfair recommendations when using the platform \cite{geyik2019fairness,lal2020fairness}. Meanwhile, researchers are investigating new approaches to incorporate fairness-aware mechanisms into RSs (i.e., fair RSs) to avoid discrimination. Early fair RSs mainly rely on statistical parity to evaluate the fairness of recommendations. For instance, demographic parity demands the same positive rate (e.g., the probability of recommending an item) for different user groups. However, recent research demonstrates that statistical parity may not be adequate to reason with fairness, as different causal relations between sensitive features and outcomes may result in divergent conclusions \cite{kusner2017counterfactual}. For example, in the Berkeley admission dataset, the lower admission rate of female applicants is because females tend to apply for difficult departments \cite{bickel1975sex}, and naively increasing the acceptance of female applicants to achieve statistical parity may be unfair to male applicants. Therefore, causality-aware fairness gains more attention, where causal models are established with domain knowledge to reason with the causal influence of sensitive features on the observed outcomes and prevent it from negatively influencing future decisions \cite{li2021towards}.

Existing causality-aware fair RSs mainly seek to eliminate all causal effects of sensitive features on recommendations, e.g., by constraining the user latent variables learned from observed ratings to be independent of sensitive features via strategies such as adversarial training \cite{wadsworth2018achieving} or maximum mean discrepancy minimization \cite{louizos2016variational}. However, a dilemma for these methods is that, most of these features may also influence user interest in a fair manner. Take race as an example. Indeed, race can be associated with various negative social stereotypes, and recommendations based on these stereotypes can be offensive to users. However, race can also determine users' cultural background \cite{schedl2018current}, such as accustomed tablewares, etc., and recommending chopsticks to East-Asian users is rarely considered offensive for online shopping platforms. Consequently, indiscriminately eliminating all the causal influence of race on recommendations may degenerate the cultural diversity critical for personalization. Another widely acknowledged example is from Pearl \cite{pearl2009causality}, which states that the education level of job applicants should not affect job recommendations based on negative stereotypes, but may indirectly influence the decision via certain job-related applicant features correlated with education level, such as skills. Therefore, a better strategy to achieve fair RS is path-specific causal analysis, where only unfair correlations between sensitive features and observed ratings are eliminated in recommendations.

However, the problem remains difficult because of the following multifaceted challenges. First, a prerequisite for most path-specific causal inference algorithms is the prior knowledge of the causal model, where factors that lead to fair or unfair correlations between sensitive features and outcomes are known and measured in advance \cite{kilbertus2017avoiding,nabi2018fair,wu2019pc,chiappa2019path}. However, this assumption does not hold for RSs, as factors that causally determine the observed user behaviors are usually latent, which makes it difficult to judge whether or not they mediate the fair influences of sensitive features and can be generalized to other users. In addition, although recent awareness of fair RS from the industry has made it possible to collect potential unfair recommendations based on users' feedback to facilitate the identification of unfair latent mediators of sensitive features, such observations are usually extremely sparse, and it is difficult to ensure fairness for users with sparse or no known unfair items (i.e., path-specific fairness for RS suffers from cold-start issues \cite{li2023transferable}). 

To address the aforementioned challenges, we propose a novel path-specific fair RS (PSF-RS) for recommendations. We first establish a causal graph to reason with the causal generation process of the biased observed ratings, assuming that the fair and unfair correlations between sensitive features and the observed ratings can be summarized into two latent proxy mediators. We then define the concept of path-specific bias (PS-Bias) based on path-specific counterfactual analysis on the causal graph, where we demonstrate that naive RSs can be unfair even if they do not explicitly use users' sensitive features for recommendations. To remedy the bias, inspired by Pearl's minimal change principle \cite{pearl2009causality}, we minimally transform the biased factual world into a hypothetically fair world with zero PS-Bias, where a fair RS model can be learned accordingly by solving a constrained optimization problem. We demonstrate that although existing fair RSs can also achieve zero PS-Bias, their modification of the biased factual world is not minimal, which destroys causal structures necessary for the diversities in recommendations. In contrast, PSF-RS eliminates the PS-Bias while maximally preserving the fair influences of sensitive features simultaneously. For the technical part, we propose a feasible implementation of PSF-RS, i.e., PSF-VAE, with weakly-supervised variational inference, where the latent proxy mediators of sensitive features can be inferred for all users with weak supervisions from the extremely sparse known unfair items. The contribution of this paper can be summarized as:

\begin{itemize}[leftmargin=0.5cm]
    \item To the best of our knowledge, we are the first to investigate path-specific fairness for RSs to ensure fairness while maximally preserving the necessary diversities in recommendations.
    \item Theoretically, a novel path-specific fair RS (PSF-RS) is proposed based on latent mediation analysis and path-specific counterfactual analysis, which minimally alters the biased factual world into a hypothetically fair world, where a fair RS can be learned accordingly by solving a constrained optimization problem.
    \item A feasible implementation of PSF-RS, i.e., PSF-VAE, is proposed based on weakly-supervised variational inference, where the fairness of recommendations can be generalized to users with sparse or no observed unfair item recommendations.
\end{itemize}

\section{Theoretical Analysis}

\subsection{Task Formulation}
The focus of this paper is on fairness of recommendations with implicit feedback \cite{hu2008collaborative}. Consider a dataset $\mathcal{D} = \{(\mathbf{r}_{i}, \mathbf{s}_{i}, \mathbf{x}_{i})\}_{i=1}^{I}$ of $I$ users, where $\mathbf{r}_{i} \in \{0, 1\}^{J}$ is a binary vector indicating whether user $i$ has interacted with each of the $J$ items, $\mathbf{s}_{i} \in \mathbb{R}^{K_{s}}$ denotes the sensitive user features such as race, gender, etc., and $\mathbf{x}_{i} \in \mathbb{R}^{K_{x}}$ denotes the non-sensitive user features that are not causally dependent on $\mathbf{s}_{i}$. Features $\mathbf{s}_{i}$ are sensitive in that carelessly basing recommendations on them may result in discrimination. In addition, due to the increasing awareness of fair RS from the industry, for a subset of users, we also collect certain items that each may consider unfair if these items are explicitly recommended (e.g., through self-reported unfair recommendations). We use another binary vector $\mathbf{r}_{b,i^{\prime}} \in \{0, 1\}^{J}$ to indicate the known unfair items for user $i^{\prime}$. $\mathbf{r}_{b,i^{\prime}}$ is extremely sparse and is unavailable for the majority of the users\footnote{In the remainder, the subscripts $i$ and $i^{\prime}$ would be omitted if no ambiguity exists. The capital non-boldface symbols $R, S, X, R_b$ are used to denote the random vectors.}. 

Observing the dilemma that sensitive features can both unfairly correlate with the observed ratings and causally influence user interests, the purpose of this paper is to design a path-specific fair RS that maximally eliminates the former while maximally preserving the latter, such that fairness can be achieved while necessary diversities in recommendations can be maximally preserved simultaneously.

\subsection{Causal Model and Assumptions}

\begin{figure}
\centering
\includegraphics[width=0.855\linewidth]{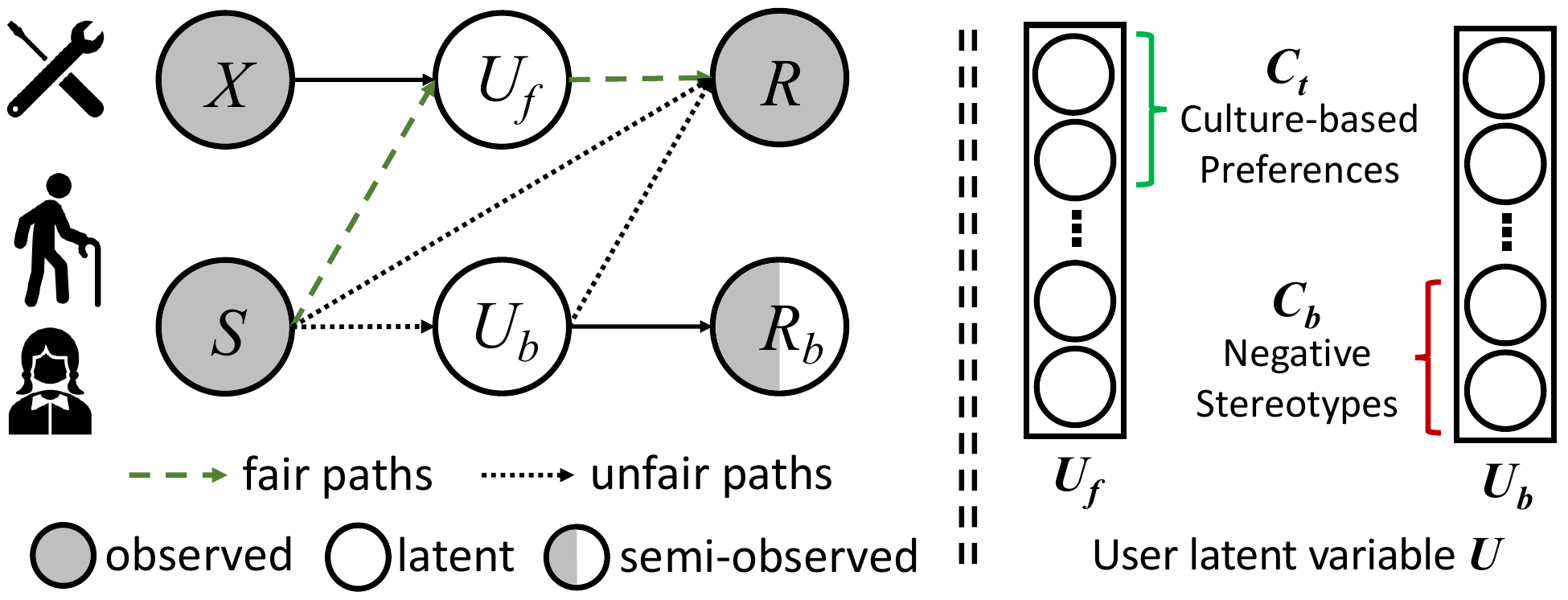}
\vspace{-2mm}
\caption{Causal graph that depicts the generation process of the observed ratings $R$ and semi-observed unfair items $R_{b}$.}
\label{fig:causal_graph}
\vspace{-2mm}
\end{figure}

Throughout this paper, we assume that the causal graph that generates the observed biased ratings $R$ and the semi-observed unfair items $R_{b}$ can be represented by Fig. \ref{fig:causal_graph}, where the edges denote the direction of causal influences. The details are introduced as follows.

\subsubsection{\textbf{User Fair Latent Variable}}

Most existing probabilistic RSs aggregate the hidden factors that causally determine the observed user behaviors $R$ into the user latent variable $U$ \cite{hu2008collaborative,li2017collaborative,liang2018variational}, which is usually assumed to be causally influenced by user features $S$ and $X$ \cite{li2021towards}. Existing fair RSs consider all the variation of $U$ due to $S$ as unfair and indiscriminately eliminate them when making new recommendations. However, we postulate that for each user, we can find $U_{f} \in \mathbb{R}^{K_{f}}$ contained in $U$ that mediates the fair influence of $S$ on $R$ (or has no causal relations with $S$). We name $U_{f}$ the user fair latent variable. $U_{f}$ has the property of being \textbf{resolving}\footnote{For readers without much  background knowledge in causal inference, we provide simple and intuitive definitions for the terms highlighted in \textbf{bold} in Appendix \ref{sec:app_causal}.} for $S$ in that any influence of $S$ on $R$ \textbf{mediated} by $U_{f}$ should be preserved to facilitate necessary diversities in recommendations. For example, sensitive feature \textit{race} can determine a user's \textit{cultural preference} $C_t$ (could be several dimensions of $U$), which is a crucial factor that determines users' personalized interest. Therefore, $C_t$ should be subsumed in $U_{f}$ such that the causal influence of $S$ on $R$ mediated by $C_t$, which can be denoted by a \textbf{causal path} $S \rightarrow C_t \rightarrow R$, is allowed to be captured by RSs to promote culture-tailored recommendations. 

\subsubsection{\textbf{User Bias Latent Variable: The Proxy Mediator}}

In addition, we use the user bias latent variable $U_{b} \in \mathbb{R}^{K_{b}}$ to summarize the remaining variations of $U$ due to $S$, which captures the unfair correlations between sensitive features $S$ and the observed ratings $R$ in the collected data. The unfair influence of $S$ mainly lies in two-fold. From the users' perspective, sensitive features $S$ can determine some social stereotypes $C_{b}$ (which could be some other dimensions of $U$) associated with certain demographic groups. Although some users may behave just according to the stereotypes (which leads to another causal path from $S$ to $R$, i.e., $S \rightarrow C_b \rightarrow R$), we should not generalize them to other users with the same sensitive features. In addition, the unfair influence of $S$ can also be attributed to the previous RS, where items unfairly associated with certain demographic groups may be overly exposed to these users that bias their behaviors \cite{liu2020general}. Formally, the assumption that describes the unfair correlations between $S$ and $R$ can be summarized as follows:
\begin{assumption}
The unfair correlations between $S$ and $R$ are composed of (1) the direct effect of $S$ on $R$; (2) all indirect mediated effects of $S$ on $R$ \textbf{not resolved} by $U_{f}$, where the latter is assumed to be able to be summarized by a one-step \textbf{latent proxy mediator} $U_{b} \in \mathbb{R}^{K_{b}}.$
\end{assumption}
The above assumption of unfair correlations between $S$ and $R$ is based on the \textit{skeptical view} of Kilbertus et al. \cite{kilbertus2017avoiding}, which states that all potential influences of sensitive features on outcomes should be assumed as discriminatory unless they can be justified by a resolving mediator, which is the user fair latent variable $U_{f}$ in our case. We summarize all indirect unfair influences of $S$ into a user bias latent variable $U_{b}$ because it is intractable to enumerate and measure all unfair mediators of sensitive features (e.g., all discriminatory stereotypes). One sufficient condition that allows such a substitution is that $U_{b}$ \textbf{blocks} every mediated unfair path between $S$ and $R$ while \textbf{unblocking} every fair path resolved by $U_{f}$. This could be the case where all unfair mediators of $S$ causally determine $U_{b}$ and through which influence $R$, which is a common assumption in latent mediation analysis \cite{albert2016causal,cheng2022causal}. Since our primary task is to analyze the fair and unfair influences of sensitive features $S$ on the observed ratings $R$, other exogenous variables that causally determine $U_{f}$ and $U_{b}$ are omitted and summarized into their uncertainties.


\begin{figure}
\centering
\includegraphics[width=0.84\linewidth]{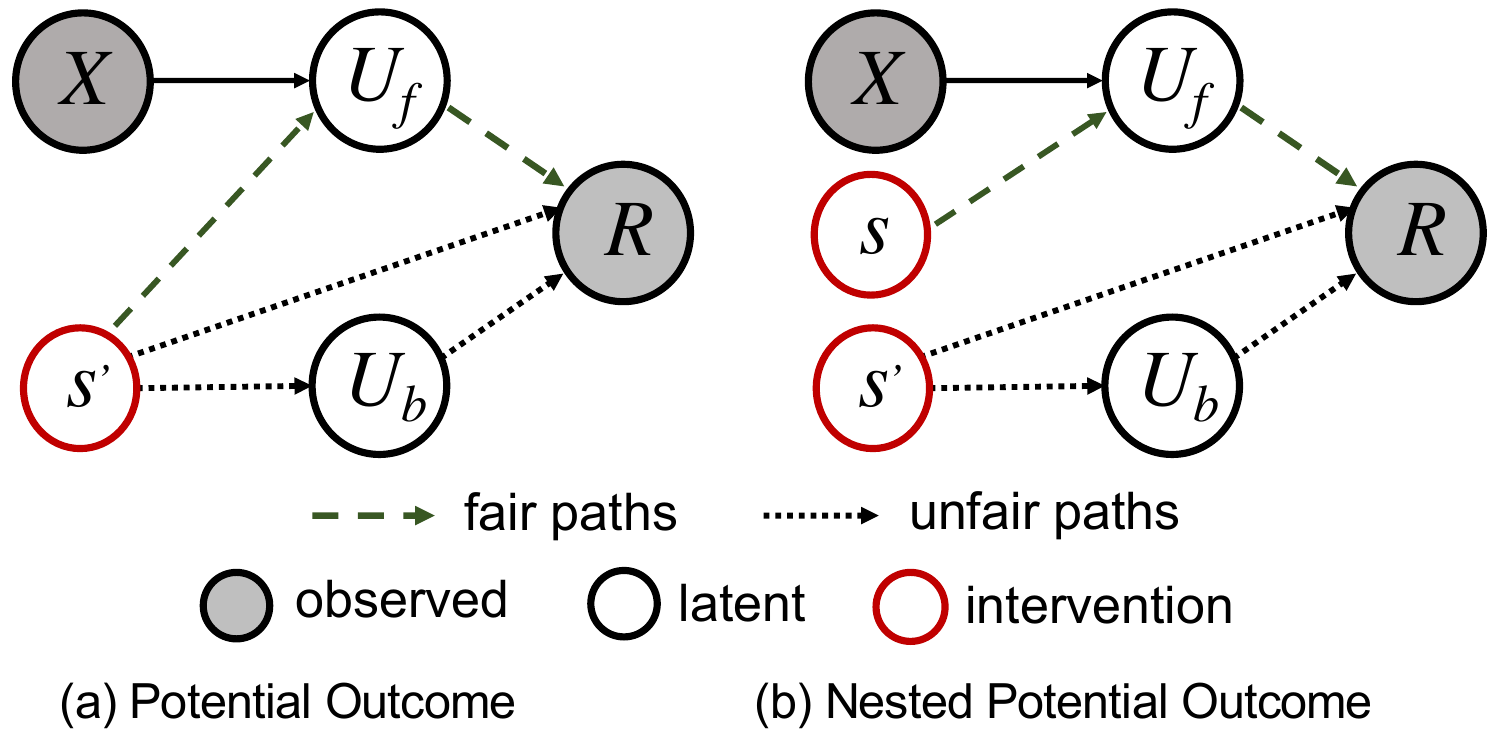}
\caption{Comparisons between potential outcome that sets sensitive features $S$ to $\mathbf{s}^{\prime}$ and nested potential outcome that sets $S$ to different values along different causal paths. }
\label{fig:npo}
\end{figure}

\subsubsection{\textbf{Path-Specific Counterfactuals}} After introducing the latent factors $U_{f}$ and $U_{b}$ that mediate the fair and unfair influences of sensitive features $S$ on observed ratings $R$ and the causal graph in Fig. \ref{fig:causal_graph}, we are ready to define the unfairness inherent in the dataset $\mathcal{D}$, which is a crucial first step toward achieving fairness in RSs. 

According to the causal graph in Fig. \ref{fig:causal_graph}, we can represent the variation of $R$ due to $S$ (with fixed $X$) in $\mathcal{D}$ with the distribution $p(R|S,X)$, which is governed by latent mediators $U_{f}$, $U_{b}$ as follows: 
\begin{equation}
p(R|S,X) = \mathbb{E}_{p(U_{f}|S,X), p(U_{b}|S)}\left[p(R|U_{f}, U_{b})\right],  
\end{equation}
where $\mathcal{F} = \{p(R|U_{f}, U_{b}), p(U_{f}|S, X), p(U_{b}|S) \}$ are the \textbf{structural equations} associated with the causal graph. However, we should note that not all variations of $R$ due to $S$ encapsulated in $p(R|S,X)$ are discriminatory, as the causal influences of $S$ mediated by $U_{f}$, e.g., the cultural-based preferences ($S \rightarrow C_{t} \rightarrow R$), are crucial manifestations of diversity and personalization in user interests. 

To address the above challenge, we measure the unfair variation of $R$ due to $S$ with path-specific counterfactual inference \cite{kusner2017counterfactual}, where we determine how ratings $R$ will change if users' sensitive features $S$ are set to a counterfactual value $\mathbf{s}^{\prime}$ along the unfair paths $S \rightarrow U_{b} \rightarrow R$ and $S \rightarrow R$, while maintaining its factual value $\mathbf{s}$ along the fair path $S \rightarrow U_{f} \rightarrow R$. To achieve this objective, it is necessary to introduce the Nested Potential Outcome (NPO) defined as follows:
\begin{definition}
\label{def:npo}
\noindent \textit{We use the \textbf{Nested Potential Outcome} (NPO) $R_{S \leftarrow \mathbf{s}^{\prime}}(U_{f, S \leftarrow \mathbf{s}}, U_{b, S \leftarrow \mathbf{s}^{\prime}})$ to denote the random variable of user ratings where user sensitive features $S$ are set to $\mathbf{s}^{\prime}$ on the unfair paths $S \rightarrow R$ and $S \rightarrow U_{b} \rightarrow R$ and to $\mathbf{s}$ on the fair path $S \rightarrow U_{f} \rightarrow R$.}
\end{definition}
\noindent The NPO $R_{S \leftarrow \mathbf{s}^{\prime}}(U_{f, S \leftarrow \mathbf{s}}, U_{b, S \leftarrow \mathbf{s}^{\prime}})$ can be intuitively represented by an intervened causal graph in Fig. \ref{fig:npo}-(b). However, the unconditional NPO reasons with the intervention conducted upon the whole population, whose factual sensitive features $S$ do not necessarily equal $\mathbf{s}$. Therefore, to constrain the NPO to users with factual sensitive feature $S=\mathbf{s}$ (and non-sensitive features $X=\mathbf{x}$) such that the fair influence of $S=\mathbf{s}$ on $R$ is excluded from the unfairness measurement, we condition it on $X=\mathbf{x}$ and $S=\mathbf{s}$ as follows:
\begin{equation}
\label{eq:hypo_user}
R_{S \leftarrow \mathbf{s}^{\prime}}(U_{f, S \leftarrow \mathbf{s}}, U_{b, S \leftarrow \mathbf{s}^{\prime}}) | X=\mathbf{x}, S=\mathbf{s}.
\end{equation}
The conditional NPO described in Eq. (\ref{eq:hypo_user}) essentially reasons with the observed ratings of hypothetical users whose sensitive features are in a "superposition" state: Their sensitive features $S$ preserve the factual value $S=\mathbf{s}$ along the fair path $S \rightarrow U_{f} \rightarrow R$ while having the counterfactual value $S=\mathbf{s}^{\prime}$ along the unfair paths $S \rightarrow U_{b} \rightarrow R$ and $S \rightarrow R$. This allows the theoretical analysis of path-specific bias/fairness of different RS models in the following subsections.

\subsection{Unfairness of Naive RSs} 
\label{sec:naive_rs}
\begin{figure}
\centering
\includegraphics[width=0.84\linewidth]{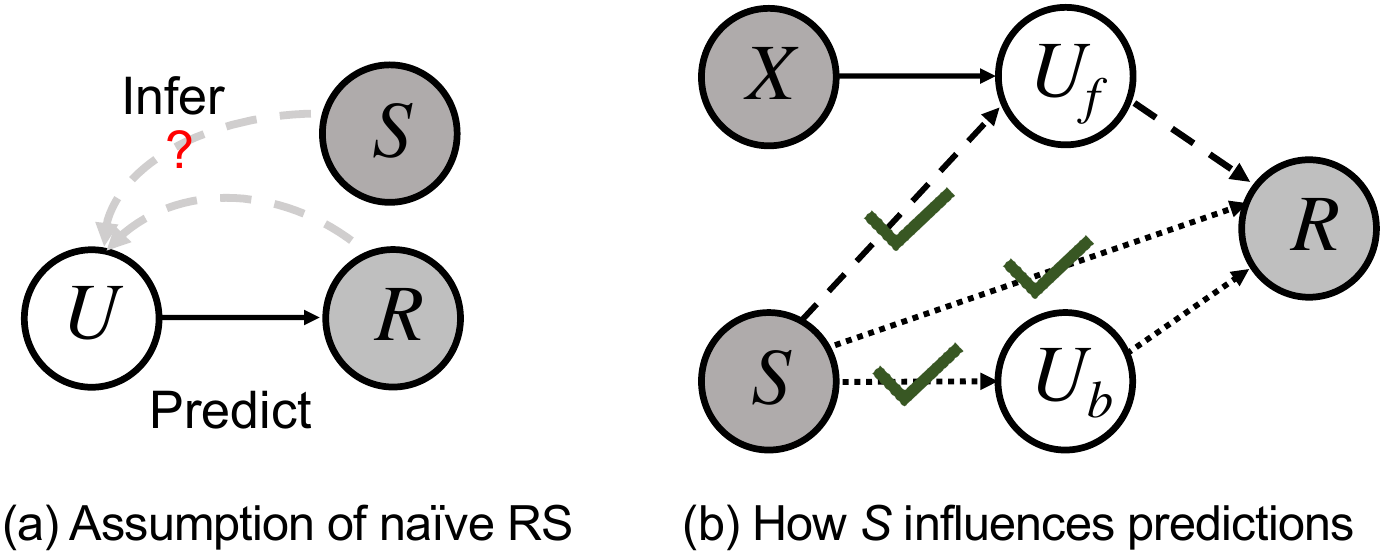}
\caption{Naive RS that infers $U$ from $R$ and (possibly) $S$ for rating predictions. If the inference is accurate, all influences of $S$ on $R$ are allowed in future recommendations.}
\label{fig:naive-rs}
\vspace{-3mm}
\end{figure}

Based on the conditional NPO, we are now ready to formally analyze the unfairness of naive RSs whose rating predictions are \textit{consistent} with the causal mechanisms that generate the biased observed ratings. We show that even if these models do not directly use sensitive features $S$ for recommendations, they can still capture the unfair correlations between $S$ and $R$ and make biased recommendations.

\subsubsection{\textbf{Path-Specific Bias for Naive RSs}} Naive RSs assume that the observed ratings $R$ are generated from user latent variables $U$ via generative distribution $p_{naive}(R|U)$\footnote{We use $p_{model}$ to represent the distributions assumed by an RS model, which should be distinguished with the structural causal equations $p$ (with no subscription) in $\mathcal{F}$ that describe the causal generative process of the biased observed ratings.}, where $p_{naive}(R|U)$ and $U$ can be obtained by maximizing the log-likelihood $\mathcal{L}$ of the observed ratings $R$ (and possibly with the support of user features $S$ and $X$) via factorization \cite{mnih2007probabilistic} or variational inference \cite{liang2018variational}. The inferred $U$ and the generative distribution $p_{naive}(R|U)$ are then used to predict new ratings for recommendations (Fig. \ref{fig:naive-rs}-(a)). If the learned generative and inference distributions of the naive RSs are accurate, $U$ captures all latent factors that causally influence the observed user behaviors $R$, i.e., $U = \{U_{f}, U_{b}\}$ (or its bijective), and $p_{naive}(R|U)$ is consistent with the causal mechanism that generates the observed ratings, i.e., $p(R|U_{f},U_{b})$. Therefore, the unfairness of the naive RSs can be quantified by the path-specific effects of $S$ on $R$ through the unfair paths on the factual causal graph, which can be defined as:
\begin{equation}
\label{eq:psb}
\begin{aligned}
 PSBias(\mathbf{x}, \mathbf{s}, \mathbf{s}^{\prime}) &= \mathbb{E}\left[R_{S \leftarrow \mathbf{s}^{\prime}}\left(U_{f, S \leftarrow \mathbf{s}},  U_{b, S \leftarrow \mathbf{s}^{\prime}}\right) \Big | X=\mathbf{x}, S=\mathbf{s}\right] \\
 &- \mathbb{E}\left[R_{S \leftarrow \mathbf{s}}\left(U_{f, S \leftarrow \mathbf{s}},  U_{b, S \leftarrow \mathbf{s}}\right) \Big| X=\mathbf{x}, S=\mathbf{s} \right].
\end{aligned}
\end{equation}
Intuitively, for users with factual features $X=\mathbf{x}$ and $S=\mathbf{s}$, path-specific bias $PSBias(\mathbf{x}, \mathbf{s}, \mathbf{s}^{\prime})$ defined in Eq. (\ref{eq:psb}) denotes the difference of rating predictions from naive RSs if their sensitive features $S$ change to $\mathbf{s}^{\prime}$ along the unfair paths $S \rightarrow R$ and $S \rightarrow U_{b} \rightarrow R$, while $S$ is held unchanged along the fair path $S \rightarrow U_{f} \rightarrow R$, and the non-sensitive features $X$ are held unchanged along all the paths. $PSBias(\mathbf{x}, \mathbf{s}, \mathbf{s}^{\prime})$ won't be zero for naive RSs if causal path $S \rightarrow U_{b} \rightarrow R$ is not trivial, but the claim is not self-evident from Eq. (\ref{eq:psb}), and we show how to calculate $PSBias(\mathbf{x}, \mathbf{s}, \mathbf{s}^{\prime})$ in the next subsection.
\begin{figure}
\centering
\includegraphics[width=0.88\linewidth]{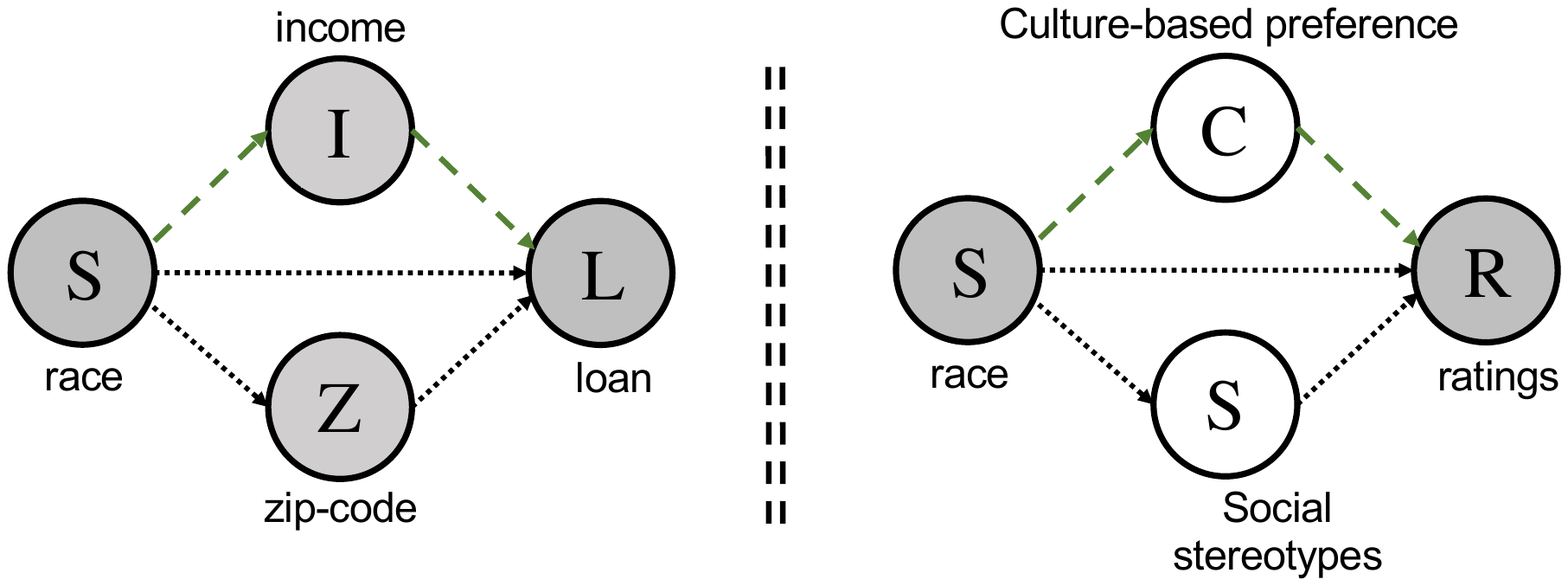}
\caption{Existing fair RS that constrains the inferred $U$ to be independent of $S$. If the constraint is satisfied, both fair and unfair influences of $S$ are blocked in recommendations.}
\vspace{-3mm}
\label{fig:fair-rs}
\end{figure}
\subsubsection{\textbf{Calculation of PS-Bias}} It is generally intractable to calculate $PSBias(\mathbf{x}, \mathbf{s}, \mathbf{s}^{\prime})$ because it contains NPOs that reason with hypothetical users with counterfactual sensitive features $S=\mathbf{s}^{\prime}$ along the unfair paths. However, with the Sequential Ignorability Assumption commonly used in causal mediation analysis \cite{imai2010general}, the first counterfactual term in Eq. (\ref{eq:psb}) can be calculated as follows:
\begin{equation}
\label{eq:bias_nested_dist}
\begin{aligned}
 \mathbb{E}&\left[R_{S \leftarrow \mathbf{s}^{\prime}}\left(U_{f, S \leftarrow \mathbf{s}},  U_{b, S \leftarrow \mathbf{s}^{\prime}}\right) \Big | X=\mathbf{x}, S=\mathbf{s} \right] \\ 
 &=\int _ {\mathbf{r}, \mathbf{u}_f, \mathbf{u}^{\prime}_b} p(\mathbf{r} | \mathbf{s}^{\prime}, \mathbf{u}_f, \mathbf{u}^{\prime}_b)  \cdot p(\mathbf{u}_f | \mathbf{s}, \mathbf{x}) \cdot p(\mathbf{u}^{\prime}_b | \mathbf{s}^{\prime}) \cdot \mathbf{r}\\
 &=\int _ {\mathbf{r}, \mathbf{u}_f, \mathbf{u}^{\prime}_b} p(\mathbf{r} | \mathbf{u}_f, \mathbf{u}^{\prime}_b(\mathbf{s}^{\prime})) \cdot p(\mathbf{u}_f | \mathbf{s}, \mathbf{x}) \cdot p(\mathbf{u}^{\prime}_b | \mathbf{s}^{\prime}) \cdot \mathbf{r}, 
\end{aligned}
\end{equation}
where in the final step, we summarize the direct unfair influence of sensitive features $S$ on ratings $R$ into $U_{b}$ for simplicity. The rigorous proof can be referred to in Appendix \ref{sec:proof_identify}. Similarly, the second factual term in Eq. (\ref{eq:psb}) can be calculated as follows:
\begin{equation}
\label{eq:nested_dist}
\begin{aligned}
 \mathbb{E}&\left[R_{S \leftarrow \mathbf{s}}\left(U_{f, S \leftarrow \mathbf{s}},  U_{b, S \leftarrow \mathbf{s}}\right)=\mathbf{r} \Big | X=\mathbf{x}, S=\mathbf{s} \right] \\ 
 &=\int _ {\mathbf{r}, \mathbf{u}_f, \mathbf{u}_b} p(\mathbf{r} | \mathbf{u}_f, \mathbf{u}_b) \cdot p(\mathbf{u}_f | \mathbf{s}, \mathbf{x}) \cdot p(\mathbf{u}_b | \mathbf{s}) \cdot \mathbf{r},
\end{aligned}
\end{equation}
where Eqs. (\ref{eq:bias_nested_dist}) and (\ref{eq:nested_dist}) can be plugged into Eq. (\ref{eq:psb}) to calculate the $PSBias(\mathbf{x}, \mathbf{s}, \mathbf{s}^{\prime})$. Clearly, $PSBias(\mathbf{x}, \mathbf{s}, \mathbf{s}^{\prime})$ for naive RSs cannot be zero, because sensitive features $S$ can unfairly influence the observed ratings $R$ via the user bias latent variable $U_{b}$, which makes the $p(U_{b}|S)$ and $p(R|U_{f}, U_{b})$ terms in Eqs. (\ref{eq:bias_nested_dist}) and (\ref{eq:nested_dist}) non-trivial.

\subsection{Minimal Change Principle and Over- Fairness of Existing Fair RSs}

To remedy the bias, existing fair RSs impose constraints upon the naive RSs. An exemplar strategy is to maximize the log-likelihood $\mathcal{L}$ of the observed ratings in $\mathcal{D}$, i.e., $\mathcal{D}_{R}$, while constraining the inferred user latent variables $U$  to be independent of the sensitive features $S$ (see Fig. \ref{fig:fair-rs}-(a)). This can be formulated as follows:
\begin{equation}
\label{eq:efairness_const}
\max_{U,\ p_{ef}} \ \mathcal{L}\left(p_{ef}(R \mid U); \mathcal{D}_{R} \right) \ s.t., U \Perp S.
\end{equation}
The constraint can be implemented via strategies such as adversarial training \cite{li2021towards} or maximum mean discrepancy (MMD) minimization \cite{louizos2016variational}. To satisfy such a  constraint, the causal mechanisms $p(U_{f}|S,X)$ and $p(U_{b}|S)$ that underlie the generation of the observed ratings must be altered into  $p_{ef}(U_{f}|X)$, $p_{ef}(U_{b})$ by dropping the dependence on $S$, which can be represented by a new causal graph illustrated in Fig. \ref{fig:fair-rs}-(b) (with causal edges marked by {\color{red}$\boldsymbol{\times}$} removed).

We can prove that existing fair RSs can eliminate the PS-Bias if the constraint is tight, such that $U$ and $S$ are strictly independent (see Appendix \ref{sec:ps-bias-generalize} and \ref{sec:nb_efrs} for details). However, it can also lead to over-fairness issues, where the causal structure $p(U_{f}| S, X)$ that denotes the fair influences of $S$ on $R$ mediated by $U_{f}$ is destroyed. Therefore, necessary diversities in recommendations due to the fair influence of sensitive features (e.g., cultural diversity) can be undesirably lost. Essentially, the independence constraint of existing fair RSs is against the Minimal Change Principle of Pearl \cite{pearl2009causality}, which states that counterfactuals (i.e., a fair rating generation model) should be reasoned with by minimally adjusting the factual world (i.e., the causal model that generates biased observed ratings).

\subsection{Path-Specific Fairness for RSs}

\begin{figure}
\centering
\includegraphics[width=0.84\linewidth]{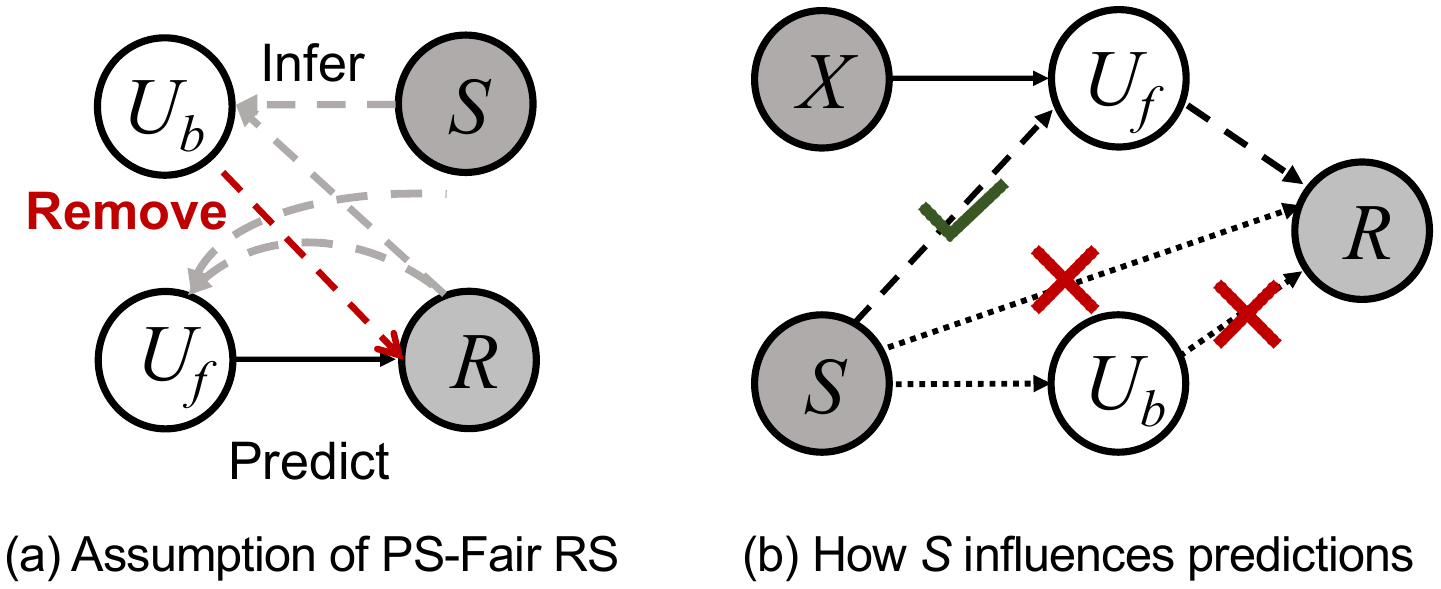}
\caption{PSF-RS that minimally changes the biased factual world represented by Fig. \ref{fig:causal_graph} into a hypothetically fair world, where a PS-Fair RS model can be learned accordingly.}
\label{fig:psf-rs}
\vspace{-3mm}
\end{figure}

To address the over-fairness drawbacks of existing fair RSs, we propose a path-specific fair RS, i.e., PSF-RS, that minimally alters the biased factual world (represented by the causal graph in Fig. \ref{fig:causal_graph}) into a hypothetically fair world, and based on it generates new ratings for recommendations. Specifically, we aim to find a counterfactual distribution $p_{psf}(R  | U_{f}, U_{b})$ close to the factual distribution $p(R | U_f, U_b)$ that causally generates the biased observed ratings (measured by KL-divergence), while inducing a new causal model with zero $PSBias^{*}(\mathbf{x},\mathbf{s},\mathbf{s}^{\prime})$\footnote{we use $^*$ to distinguish the PS-Bias of new causal model induced by PSF-RS from the PS-Bias of naive RSs that recommend according to the biased factual causal model.}, where other factual causal mechanisms in $\mathcal{F}$, i.e., $p(U_{f}|S,X)$ and $p(U_{b}|S)$, remain unchanged.

Assuming for now that the latent mediators $U_{f}$ and $U_{b}$ are known for each user (where the inference of $U_{f}$ and $U_{b}$ with weak supervision in $R_{b}$ will be thoroughly discussed in the next section), since the observed ratings $R$ in the dataset $\mathcal{D}$ are generated according to $p(R | U_f, U_b)$, the minimization of the KL between $p_{psf}(R  | U_{f}, U_{b})$ and $p(R | U_f, U_b)$ is equivalent to the maximization of the likelihood $\mathcal{L}$ of the observed ratings in $\mathcal{D}$. Therefore, the objective of PSF-RS can be formulated as a constrained optimization problem as follows:
\begin{equation}
\label{eq:fairness_const}
\max_{p_{psf}} \ \mathcal{L}\left(p_{psf}(R \mid U_{f}, U_{b}); \mathcal{D}_{R}\right) \ s.t., \ PSBias^{*}(\mathbf{x},\mathbf{s},\mathbf{s}^{\prime})=0,\  \forall \mathbf{x},\mathbf{s},\mathbf{s}^{\prime}. 
\end{equation}
The constraint essentially restricts the family of RS models that we can use for recommendations into the ones that induce a new causal model with zero $PSBias^{*}(\mathbf{x},\mathbf{s},\mathbf{s}^{\prime})$. The simplest distribution family that satisfies the constraint is the one that uses only $U_{f}$ to generate recommendations, i.e., $p_{psf}(R \mid U_{f})$ (see Appendix \ref{sec:proof_zero_bias} for the proof of zero PS-Bias for the PSF-RS). The newly-induced causal graph that changes $p(R|U_{f}, U_{b})$ to $p_{psf}(R|U_{f})$ while keeping $p(U_{f}|S,X)$ and $p(U_{b}|S)$ intact is shown in Fig. \ref{fig:psf-rs}-(b) for reference.

\section{PS-FAIR VARIATIONAL AUTO-ENCODER}

Previous sections have demonstrated PSF-RS's theoretical advantage of achieving path-specific fairness while maximally preserving the necessary diversities in recommendations. However, its practical implementation still faces two challenges as follows:
\begin{itemize}[leftmargin=0.5cm]
    \item First, since both fair and unfair mediators of $S$, i.e., $U_{f}$ and $U_{b}$, are latent, the objective of PSF-RS in Eq. (\ref{eq:fairness_const}) cannot be directly optimized to obtain the PS-Fair rating predictor $p_{psf}(R | U_{f})$.
    \item  In addition, although the known unfair items $R_{b}$, i.e., another indirect causal effect of $S$ mediated by $U_{b}$, can be used to infer $U_{b}$ and distinguish it from $U_{f}$, $R_{b}$ is extremely sparse and is only partially observable for a small subset of users. 
\end{itemize}
To address the aforementioned challenges, we propose a novel semi-supervised deep generative model called path-specific fair variational auto-encoder (PSF-VAE) as the implementation of PSF-RS. Specifically, in the \textit{factual} modeling step, PSF-VAE infers $U_{f}$ and $U_{b}$ from the biased observational ratings $R$ in the dataset $\mathcal{D}$ via deep neural networks (DNNs), where user features $S$ and $X$ are used as extra covariates and $R_{b}$ as additional weak supervision signals. Then, in the \textit{counterfactual} reasoning step, $U_{b}$ that explains away the unfair influences of $S$ is eliminated according to Eq. (\ref{eq:fairness_const}), and $U_{f}$ that maximally preserves the fair influence of $S$ and other aspects of user interests is utilized to generate new recommendations.

\subsection{Factual Generative Process}
The factual generative process of PSF-VAE is consistent with the causal model in Fig. \ref{fig:causal_graph}, such that latent mediators $U_{f}$ and $U_{b}$ can be properly inferred from the biased observational data. PSF-VAE starts by generating for each user the user fair and bias latent mediators $U_{f}$ and $U_{b}$ from Gaussian priors $p_{\theta}(U_{f}|S,X)$ and $p_{\theta}(U_{b}|S)$ as 
\begin{equation}
 \mathbf{u}_{f} \sim \mathcal{N}(f_{uf}([\mathbf{s} || \mathbf{x}]), \mathbf{I}_{K_{f}}), \ \mathbf{u}_{b} \sim \mathcal{N}(f_{ub}(\mathbf{s}), \mathbf{I}_{K_{b}}),   
\end{equation}
where $f_{uf}$ and $f_{ub}$ are two functions, $[\cdot || \cdot]$ represents vector concatenation, and $\theta$ denotes the trainable parameters associated with the generative network, respectively. Then, for the small subset of users with known unfair items $\mathbf{r}_{b}$, $\mathbf{r}_{b}$ are generated from $\mathbf{u}_{b}$ via $p_{\theta}(R_{b}|U_{b})$ parameterized as the following Bernoulli distribution, 
\begin{equation}
\label{eq:dvae}
\mathbf{r}_{b} \sim Bernoulli(MLP_{b}(\mathbf{u}_{b})),
\end{equation}
where $MLP_{b}$ is a multi-layer perceptron (MLP) with sigmoid final layer activation ($sigmoid(\textbf{x}) = 1/(1+e^{-\textbf{x}})$). Finally, the observed ratings $\mathbf{r}$ are generated from both $\mathbf{u}_{f}$ and $\mathbf{u}_{b}$ via $p_{\theta}(R|U_{f}, U_{b})$ parameterized as the following multinomial distribution,
\begin{equation}
\label{eq:biased_ratings}
\mathbf{r} \sim Multi(MLP_{r}([\mathbf{u}_{f}  || \mathbf{u}_{b} ]), N),\end{equation} 
where $MLP_{r}$ is another MLP with softmax final layer activation, i.e., $[softmax(\mathbf{x})]_{i}=e^{x_{i}}/{\sum_{j}e^{x_{j}}}$; $N$ is the number of interacted items. 
 
\subsection{Weakly-Supervised Variational Inference}

Given that the (factual) generative distributions of both $R$ and $R_{b}$ are parameterized by DNNs, and $R_{b}$ is only partially observable for a small subset of users, the true posterior distributions of the latent variables, i.e., $p_{\theta}(U_{f}|R,S,X)$ and $p_{\theta}(U_{b}|R_{b},R,S)$, are intractable. Therefore, we resort to variational inference \cite{blei2017variational,liang2018variational}, where we introduce tractable distribution families of $U_{f}$ and $U_{b}$ parameterized by DNNs with trainable parameters $\phi$, i.e., $q_{\phi}(U_{f}|\cdot)$ and $q_{\phi}(U_{b}|\cdot)$, and in $q_{\phi}$ find the distributions closest to the true but intractable posteriors measured by KL-divergence as the approximations. 

The variational posterior for $U_{f}$, i.e., $q_{\phi}(U_{f} | R, S, X)$, is straightforward. However, for $U_{b}$, we eschew the normally-adopted variational posterior $q_{\phi}(U_{b} | R_{b}, R, S)$ but use $q_{\phi}(U_{b} | R, S)$ with $R_{b}$ omitted instead, such that the inference of $U_{b}$ does not depend on the partially observed $R_{b}$. Therefore, it can be generalized to users with no observed unfair items. Under such circumstances, if $R$ and $S$ contain sufficient information of $R_{b}$, which can be guaranteed since both $R$ and $R_{b}$ are under the unfair causal influence of $S$ mediated by $U_{b}$, weak supervision signals in $R_{b}$ from the subset of users with observed unfair items can still guide the training of the inference network $q_{\phi}(U_{b} | R, S)$ to provide good variational approximations.

\subsection{Evidence Lower Bound}
The minimization of the KL-divergence between variational and true posterior distributions is equivalent to the maximization of the evidence lower bound (ELBO) as (proofs see Appendix \ref{sec:elbo_proof})\footnote{In practice, we further simplify the ELBO by dropping the dependence of the priors of $U_{f}$ and $U_{b}$ on $S$ and $X$, i.e., $\mathbf{u}_{f} \sim \mathcal{N}(\mathbf{0}, \mathbf{I}_{K_{f}}), \ \mathbf{u}_{b} \sim \mathcal{N}(\mathbf{0}, \mathbf{I}_{K_{b}})$. In addition, we first optimize the $U_{b}$-specific terms in the ELBO, and then fix $U_{b}$ and learn other terms.}:
\begin{equation}
\label{eq:elbo}
\begin{aligned}
    & \ \ \ ELBO = \mathbb{E}_{q_{\phi}(U_{f}, U_{b} | \cdot)} [\ln p_{\theta}(R | U_{f}, U_{b})]  + \mathbb{E}_{q_{\phi}(U_{b} | \cdot)}[p_{\theta}(R_{b} | U_{b})] \\
    &- \mathbb{KL}[q_{\phi}(U_{f} | R, S, X) || p_{\theta}(U_{f}|S,X)]- \mathbb{KL}[q_{\phi}(U_{b} | R, S) \ || \ p_{\theta}(U_{b}|S)],
\end{aligned}
\end{equation}
\noindent which is a lower bound of the model evidence $\ln p_{\theta}(R, R_{b}|S,X)$. In Eq. (\ref{eq:elbo}), the first two terms are the expected log-likelihood of $R$ and $R_{b}$ given the latent mediators $U_f$ and $U_b$, which encourage $U_f$ and $U_b$ to best explain the observed biased ratings (where the bias in $R$ is explained-away from $U_{f}$ by $U_{b}$), and the last two terms are the KL-divergence between the variational posteriors and the priors. 

For users with no observed unfair items $R_{b}$, the second expected log-likelihood term $\mathbb{E}_{q_{\phi}(U_{b} | R, S)}[p_{\theta}(R_{b} | U_{b})]$ is dropped from the ELBO, and we only use the observed ratings $R$ and the user sensitive features $S$ to infer the corresponding user bias latent variable $U_{b}$ via the variational posterior $q_{\phi}(U_{b} | R, S)$. For these users, when maximizing the first term of the ELBO, i.e., $\mathbb{E}_{q_{\phi}(U_{f}, U_{b} | \cdot)} [\ln p_{\theta}(R | U_{f}, U_{b})]$, the inferred $U_{b}$ \textbf{can still help explain away} the unfair influence of $S$ on $R$, such that $U_{f}$ can focus exclusively on capturing the fair user interests that are generalizable to future recommendations.

\subsection{Disentanglement via Adversarial Training}

Before introducing $p_{psf}(R | U_{f})$ that minimally changes the biased factual world into a hypothetically fair world to make fair recommendations, we note that the theoretical PS-Fairness of PSF-RS requires a correctly specified inference model (as Eq. (\ref{eq:fairness_const}) requires known $U_{f}$ and $U_{b}$). Especially, we need to ensure $U_{f} \Perp U_{b} | S$, which prevents $U_{f}$ from \textbf{directly} depending on $U_{b}$, such that the unfair information of $S$ cannot be leaked to $U_{f}$. Since the true posteriors of $U_{f}$ and $U_{b}$ are not guaranteed to be in the variational family $q_{\phi}$, the unfair information of $S$ in $U_{b}$ may be leaked to $U_{f}$ due to potential mis-specification of the inference model, especially when supervision signals in $R_{b}$ are available only for a subset of users. 

We utilize an adversarial training-based strategy \cite{goodfellow2020generative} to ensure the conditional independence of $U_{f}$ and $U_{b}$ given $S$ in case of inference model mis-specification. Following \cite{bellot2019conditional}, we first parameterize a discriminator model $p_{d}$ that predicts $U_{b}$ from $U_{f}$ and $S$ as:
\begin{equation}
p_{d}(U_{b} | U_{f}, S) = \mathcal{N}(MLP_{d}([U_{f}||S]), \mathbf{I}_{K_{d}}).
\end{equation}
Then, concurrent with the maximization of the ELBO in Eq. (\ref{eq:elbo}), $U_{f}$ and $U_{b}$ obtained from variational posteriors $q_{\phi}$ are used to train the discriminator $p_{d}$. Specifically, we fix $\hat{q}_{\phi}(U_{b} | R, S)$, sample $\hat{\mathbf{u}}_{b}$ from it and train the discriminator $p_{d}(U_{b} | U_{f}, S)$ to best predict $\hat{\mathbf{u}}_{b}$ from $U_{f}$ and $S$. Meanwhile, we constrain the inference model of $U_{f}$, i.e., $q_{\phi}(U_{f} | R,S,X)$, to fool the discriminator. The above process can be formulated as a GAN-like mini-max game as follows:
\begin{equation}
\label{eq:minimax}
   \min_{q_{\phi}} \max_{p_{d}} \mathbb{E}_{q_{\phi}(U_{f} | R,S,X)}[\ln p_{d}(\hat{\mathbf{u}}_{b}  | U_{f}, S)], \  \hat{\mathbf{u}}_{b} \sim \hat{q}_{\phi}(U_{b} | R,S).
\end{equation}
With a sufficient capacity of the discriminator $p_{d}$, Li et al. \cite{li2021towards} showed that $U_{f} \Perp U_{b} | S$ holds when the equilibrium of Eq. (\ref{eq:minimax}) is achieved. Therefore, the direct dependence of $U_{f}$ on $U_{b}$ that leads to the leak of unfair information of $S$ can be further mitigated. 

\subsection{PS-Fair Rating Predictions}

Finally, we introduce $p_{psf}(R | U_{f})$, the counterfactual rating generator that minimally modifies the biased factual world while ensuring path-specific fairness and necessary diversities in recommendations. Specifically, after optimizing the "factual step" of PSF-VAE via Eqs. (\ref{eq:elbo}) and (\ref{eq:minimax}), we fix $q_{\phi}(U_{f}|R,S,X)$ and obtain the user fair latent variables $\hat{\mathbf{u}}_{f}$ as the posterior mean. Then the PS-Fair rating predictor $p_{psf}(R | U_{f})$ can be obtained by optimizing Eq. (\ref{eq:fairness_const}) with the inferred $\hat{\mathbf{u}}_{f}$ and the observed ratings $\mathbf{r}$. Specifically, we parameterize $p_{psf}(R | U_{f})$ as the following multinomial distribution,
\begin{equation}
\label{eq:multi_pred}
\mathbf{r} \sim Multi(MLP_{psf}(\hat{\mathbf{u}}_{f}  ), N),
\end{equation}
where $MLP_{psf}$ is another MLP with softmax as the last layer activation. Finally, the multinomial probabilities of all previously uninteracted items can be obtained via $p_{psf}(R | U_{f})$, which are then ranked such that $M$ most relevant ones are fetched for recommendations.

\section{Experiments}

In this section, we present the extensive experiments conducted on two semi-simulated datasets and one real-world dataset to demonstrate the effectiveness of the proposed PSF-VAE, with an emphasis on answering the following three research questions\footnote{Codes are available at \url{https://github.com/yaochenzhu/PSF-VAE}.}:
\begin{itemize}[leftmargin=0.5cm]
    \item \textbf{RQ1.} How well can PSF-VAE achieve fairness compared with different RS methods with and without fairness constraints?
    \item \textbf{RQ2.} How well can PSF-VAE preserve necessary fair influences of sensitive features compared with existing fair RS algorithms?
    \item \textbf{RQ3.} How does the number of users with known unfair items $R_{b}$ influences the fairness performance of PSF-VAE? 
\end{itemize}

\subsection{Datasets}

It is difficult to directly evaluate PSF-VAE on real-world datasets, as the true fair and unfair causal effects of sensitive features on the observed ratings cannot be identified from the datasets. Therefore, we first establish semi-simulated datasets with known causal mechanisms between sensitive features and rating observations. We then introduce a real-world dataset collected from LinkedIn\footnote{\url{https://www.linkedin.com/}.}, where for a subset of users, their negative feedback on recommendations (i.e., explicit dismissals of Ads) is treated as the proxy of unfair items.

\begin{table}[t]
  \caption{Statistics of the semi-simulated (ML-1M and AM-VG) and the real-world (LinkedIn) datasets. \#Int. stands for the number of observed interactions. Sps. ($R$) and Sps. ($R_{b}$) denote the sparsity of observed ratings, unfair items, respectively.}
  \label{tab:datasets}
 \vspace{-3mm}
  \begin{tabular}{lccccc}
    \toprule
    Dataset & \#Int. & \#Users & \#Items & Sps. ($R$) & Sps. ($R_{b}$)\\
    \midrule
    ML-1M   & 993,504   & $6,000$ & $3,706$ &  95.53\%& 99.76\% \\
    AM-VG   & 127,741   & $7,253$ & $4,338$ & 99.60\% & 99.93\%\\
    LinkedIn & 1,055,241 & $8,896$ & $5,931$ & 98.01\% & 99.62\%\\
  \bottomrule
\end{tabular}
\end{table}

\subsubsection{\textbf{Semi-Simulated Dataset}} 
\label{sec:semi-simulated}
The semi-simulated datasets are established based on the widely-used MovieLens-1M (ML-1M) \cite{harper2015movielens} and Amazon Videogames (AM-VG) datasets \cite{mcauley2015image}. For each dataset, we train a Multi-VAE model \cite{liang2018variational} on the binarized ratings, where the decoder $f_{gen} = MLP_{gen}(\mathbf{u})$ maps the user latent variable $U \sim \mathcal{N}(\mathbf{0}, \mathbf{I}_{K})$ to the multinomial parameters $\tilde{R}$ of the ratings $R$. The latent dimension $K$ is fixed to 200 as \cite{liang2018variational}. We then assume that the first $K_{f}$ and the remaining $K_{b} = K-K_{f}$ dimensions of $U$, which we denote as $U_{f}$ and $U_{b}$, mediate the fair and unfair influences of sensitive features $S$ on the observed ratings $R$, respectively. In the simulation, for each user, we first generate a confounder $\mathbf{c} \sim \mathcal{N}(\mathbf{0}, \mathbf{I}_{K_{f}})$ that simultaneously affects $\mathbf{u}_{f}$ and $\mathbf{u}_{b}$, where user sensitive features $\mathbf{s}$ are derived from $\mathbf{c}$ by $PCA(\mathbf{c}, K_{s})$. The fair and unfair latent mediators $\mathbf{u}_{f}$ and $\mathbf{u}_{b}$ are then generated as follows:
\begin{equation*}
\begin{aligned}
\mathbf{u}_{f} = \lambda _ {f} \mathbf{c} + \sqrt{(1-\lambda^{2} _ {f})} \boldsymbol{\epsilon}_{f}; \ \  \mathbf{u}_{b}  =  \lambda _ {b} Redim(\mathbf{c}, K_{b}) + \sqrt{(1-\lambda^{2} _ {b})} \boldsymbol{\epsilon}_{b},
\end{aligned}
\end{equation*}
where the exogenous variables $\boldsymbol{\epsilon}_{f} \sim \mathcal{N}(\mathbf{0}, \mathbf{I}_{K_{f}})$, $\boldsymbol{\epsilon}_{b} \sim \mathcal{N}(\mathbf{0}, \mathbf{I}_{K_{b}})$, the function $Redim$ reduces the dimension of $\mathbf{c}$ to $K_{b}$ through random selection, and the coefficients $\lambda_{f}$ and $\lambda_{b}$ determine the noise level of $\mathbf{u}_{f}$ and $\mathbf{u}_{b}$, which are empirically fixed as 0.9 and 0.9, respectively. 

The observed ratings are generated from $\mathbf{u}_{f}$ and $\mathbf{u}_{b}$ by first calculating the multinomial parameters $\tilde{\mathbf{r}} = f_{gen}([\mathbf{u}_{f} || \mathbf{u}_{b}])$, where the top $100 \times p_{r}\%$ (ranked among all users) are selected as the rating observations $\mathbf{r}$. $p_{r}$ is set to be the same as the original datasets. The unfair items $\mathbf{r}_{b}$ are simulated with the sub-network $f^{b}_{gen}$ in $f_{gen}$ that corresponds to $\mathbf{u}_{b}$\footnote{If we denote $f_{gen}(\mathbf{u})$ as $\tilde{f}_{gen}(\mathbf{W}\mathbf{u}+\mathbf{b})$, the subnetwork can be obtained by $f^{b}_{gen} = \tilde{f}_{gen}(\mathbf{W}_{:,K-K_{b}:K}\mathbf{u}_{b}+\mathbf{b})$, where $\mathbf{W}_{:,K-K_{b}:K}$ selects the last $K_{b}$ columns of $\mathbf{W}$.}. Similarly, we obtain the multinomial parameters $\tilde{\mathbf{r}}_{b} = f^{b}_{gen}(\mathbf{u}_{b})$, where the top $100 \times p_{b}\%$ are selected as the unfair items. $p_{b}$ is determined such that the ratio of the average number of observed ratings and unfair items is the same as the real-world dataset introduced later. We do not simulate non-sensitive features $\mathbf{x}$ because the sequential ignorability assumption automatically holds with the above data generation process.

\subsubsection{\textbf{Real-World Dataset}}

In addition, we collect a real-world dataset from LinkedIn for job recommendations, where ratings $R$ denote users' interactions with the job Ads. We use the data where users actively dismissed the recommended jobs as substitutes for the unfair items $R_{b}$. User sensitive features $S$ include age, gender, and education level, all of which can influence the job recommendation in a fair manner. For example, age can determine the experience and seniority of the users, whereas education level can determine their knowledge and skills. To avoid privacy issues in user data collection, we train a generative model (VAE) to encode the raw data into a joint distribution $p_{gen}(R, R_{b}, S)$ where $S$ is embedded into a $50$-dimensional continuous vector, and we generate anonymized data from $p_{gen}$ accordingly for the experiments to protect privacy \cite{zhang2018differentially}. The statistics of the datasets are summarized in Table \ref{tab:datasets}.

\subsection{Experimental Settings}

\subsubsection{\textbf{Setups}} In our experiments, we randomly split the users into train, validation, and test sets based on the ratio of 8:1:1 \cite{liang2018variational}. For each user, 20\% of the observed ratings are held out for evaluation. For the ML-1M and AM-VG datasets, the simulated unfair items $\mathbf{r}_{b}$ for $100\times(1-c_{r})\%$ of the training and validation users are masked out as zero (where $c_{r}$ is set to $0.3$ as with the LinkedIn dataset), while $\mathbf{r}_{b}$ for all test users are used to obtain unbiased evaluations of the fairness of different methods. In our experiments, we first fix the simulated dimension of $U_{b}$, i.e., $K_{b}$, to 50 in the ML-1M and AM-VG datasets to compare the recommendation performance and fairness across different methods. We then simulate the datasets with varied $K_{b}$ to further demonstrate the robustness of PSF-VAE to different levels of unfair correlations between observed ratings and sensitive features. Finally, we show the sensitivity of PSF-VAE to the percentage of users with observed unfair items. All reported results are averaged over ten random splits of the datasets. 

\subsubsection{\textbf{Evaluation Metrics}} We evaluate different RSs from two aspects: \textit{recommendation performance} and \textit{fairness}. The recommendation performance is measured by two widely-used ranking-based metrics: Recall (R@$M$) and truncated normalized discounted cumulative gain (N@$M$)\footnote{We also use the recommendation quality (i.e., R@$M$ and N@$M$) as an indirect measure of RSs' ability to preserve the fair influences of sensitive features on ratings.}. Fairness is measured by the hit rate of top $M$ items on unfair items (HiR@$M$). For the semi-simulated datasets, the true unfair items $\mathcal{J}_{b,i}$ are available for all test users, while for the LinkedIn dataset, we can only calculate $\operatorname{HiR} @ M$ for test users with observed unfair items. In our experiments, we find that $M$ generally does not affect the relative performance of different methods. Therefore, we set $M$ to 20 for Recall and $100$ for NDCG as with \cite{liang2018variational}, and set $M$ to 10 for HiR due to the sparsity of observed unfair items.

\subsubsection{\textbf{Model Selection}}
\label{sec:model_sel}
During the training stage, we monitor the composite metric $Met_{rf}(i)$ = R@20($i$) + N@100($i$) - HiR@10($i$) on validation users with known unfair items and $Met_{r}(i)$ = R@20($i$) + N@100($i$) on validation users with no observed unfair items, and calculate the weighted average of $Met_{rf}$ and $Met_{r}$, i.e., $\hat{Met}$, over all validation users. We then select the model with the largest $\hat{Met}$ and report the recommendation and fairness metrics on test users.

\begin{table}[]
\setlength{\tabcolsep}{2pt}
\vspace{-3mm}
\centering
\caption{Comparison between PSF-VAE and various baselines. $\uparrow$ denotes the larger the better, while $\downarrow$ denotes the opposite.}
\vspace{-2mm}
\label{tab:com_results}
\small
\begin{tabular}{lccc}
\toprule
{\textbf{AM-VG}} & \textbf{Rec: R@20} $\uparrow$ & \textbf{Rec: N@100} $\uparrow$ & \textbf{Fair: HiR@10}  $\downarrow$   \\ \midrule

Multi-VAE & 0.2454 {\scriptsize $\pm$ 0.0130} & 0.2350 {\scriptsize $\pm$ 0.0093} & 0.0297 {\scriptsize $\pm$ 0.0030} \\ \midrule
CondVAE   & \textbf{0.2780} {\scriptsize $\pm$ 0.0103} & \textbf{0.2599} {\scriptsize $\pm$ 0.0058} & 0.0315 {\scriptsize $\pm$ 0.0045} \\ 
CondVAE-ES & 0.2686 {\scriptsize $\pm$ 0.0115} & 0.2493 {\scriptsize $\pm$ 0.0061} & 0.0302 {\scriptsize $\pm$ 0.0053} \\ \midrule
Fair-MMD & 0.2304 {\scriptsize $\pm$ 0.0118} & 0.2147 {\scriptsize $\pm$ 0.0094} & 0.0279 {\scriptsize $\pm$ 0.0025} \\ 
Fair-ADV & 0.2285 {\scriptsize $\pm$ 0.0081} & 0.2119 {\scriptsize $\pm$ 0.0076} & \textbf{0.0274} {\scriptsize $\pm$ 0.0020} \\ \midrule
PSF-NN & \underline{0.2702} {\scriptsize $\pm$ 0.0124} & \underline{0.2549} {\scriptsize $\pm$ 0.0095} & 0.0310 {\scriptsize $\pm$ 0.0029} \\
PSF-VAE  & 0.2691 {\scriptsize $\pm$ 0.0104} & 0.2507 {\scriptsize$\pm$ 0.0075} & \underline{0.0288} {\scriptsize$\pm$ 0.0032} \\ \bottomrule
\specialrule{0em}{-2pt}{-2pt}
 \\ \toprule 
{\textbf{ML-1M}} & \textbf{Rec: R@20} $\uparrow$ & \textbf{Rec: N@100} $\uparrow$ & \textbf{Fair: HiR@10}  $\downarrow$   \\ \midrule
Multi-VAE & 0.5493 {\scriptsize $\pm$ 0.0133} & 0.6556 {\scriptsize $\pm$ 0.0064} & 0.0938 {\scriptsize $\pm$ 0.0075} \\ \midrule
CondVAE   & \textbf{0.5689} {\scriptsize $\pm$ 0.0145} & \textbf{0.6757} {\scriptsize $\pm$ 0.0065} & 0.0953 {\scriptsize $\pm$ 0.0077} \\ 
CondVAE-ES & 0.5615 {\scriptsize $\pm$ 0.0151} & 0.6665 {\scriptsize $\pm$ 0.0069} & 0.0949 {\scriptsize $\pm$ 0.0080} \\ \midrule
Fair-MMD & 0.5312 {\scriptsize $\pm$ 0.0119} & 0.6350 {\scriptsize $\pm$ 0.0069} & 0.0893 {\scriptsize $\pm$ 0.0074} \\ 
Fair-ADV & 0.5304 {\scriptsize $\pm$ 0.0129} & 0.6348 {\scriptsize $\pm$ 0.0060} & \textbf{0.0886} {\scriptsize $\pm$ 0.0063} \\ \midrule
PSF-NN & \underline{0.5654} {\scriptsize $\pm$ 0.0104} & \underline{0.6701} {\scriptsize $\pm$ 0.0051} & 0.0942 {\scriptsize $\pm$ 0.0040} \\
PSF-VAE  & 0.5601 {\scriptsize $\pm$ 0.0148} & 0.6668 {\scriptsize$\pm$ 0.0070} & \underline{0.0904} {\scriptsize$\pm$ 0.0084} \\ \bottomrule
\specialrule{0em}{-2pt}{-2pt}
 \\ \toprule 
{\textbf{LinkedIn}} & \textbf{Rec: R@20} $\uparrow$ & \textbf{Rec: N@100} $\uparrow$ & \textbf{Fair: HiR@10}  $\downarrow$   \\ \midrule
Multi-VAE & 0.1665 {\scriptsize $\pm$ 0.0043} & 0.2553 {\scriptsize $\pm$ 0.0046} & 0.0703 {\scriptsize $\pm$ 0.0034} \\ \midrule
CondVAE   & \textbf{0.2056} {\scriptsize $\pm$ 0.0037} & \textbf{0.3042} {\scriptsize $\pm$ 0.0031} & 0.0718 {\scriptsize $\pm$ 0.0037} \\ 
CondVAE-ES & 0.1991 {\scriptsize $\pm$ 0.0047} & 0.2965 {\scriptsize $\pm$ 0.0036} & 0.0705 {\scriptsize $\pm$ 0.0023} \\ \midrule
Fair-MMD & 0.1579 {\scriptsize $\pm$ 0.0054} & 0.2398 {\scriptsize $\pm$ 0.0066} & 0.0608 {\scriptsize $\pm$ 0.0040} \\
Fair-ADV & 0.1573 {\scriptsize $\pm$ 0.0062} & 0.2372 {\scriptsize $\pm$ 0.0070} & \textbf{0.0591} {\scriptsize $\pm$ 0.0034} \\  \midrule
PSF-NN & \underline{0.2032} {\scriptsize $\pm$ 0.0024} & \underline{0.3005} {\scriptsize $\pm$ 0.0028} & 0.0709 {\scriptsize $\pm$ 0.0023} \\
PSF-VAE  & 0.2024 {\scriptsize $\pm$ 0.0045} & 0.2987 {\scriptsize$\pm$ 0.0034} & \underline{0.0647} {\scriptsize$\pm$ 0.0029} \\ \bottomrule
\end{tabular}
\vspace{-5mm}
\end{table}

\subsection{Comparisons with Baselines}
\subsubsection{\textbf{Baseline Descriptions}} To answer \textbf{RQs 1} and \textbf{2}, we compare the proposed PSF-VAE with various state-of-the-art RSs with/ without fairness-aware mechanisms. The main baselines included for comparisons can be categorized into four classes as follows:
\begin{itemize}[leftmargin=0.5cm]
    \item \textbf{Unawareness.}  RSs with unawareness use only seemingly non-sensitive information (i.e., observed ratings and non-sensitive features) for recommendations. In this regard, the Unawareness counterpart of PSF-VAE is the vanilla \textbf{Multi-VAE} \cite{liang2018variational}.
    \item \textbf{Naive.} Naive RSs explicitly utilize the sensitive features $S$ for recommendations. In our case, it can be implemented as a generalized Multi-VAE where the rating inputs are augmented with the sensitive features $S$. The augmentation is implemented as with the user conditional Multi-VAE (\textbf{CondVAE}) in \cite{pang2019novel}.
    \item \textbf{Total Fairness.} RSs with total fairness block all the effects of sensitive features $S$ on recommendations. Built upon the Unawareness model (i.e., Multi-VAE), the inferred user latent variables $U$ are constrained to be disentangled from the user sensitive features $S$ while fitting on the observed ratings $R$. We consider the following two disentanglement strategies:
    \begin{itemize}[leftmargin=0.4cm]
        \item \textbf{Fair-ADV}. Fair-ADV constrains the user latent variables of Multi-VAE to be independent with sensitive features $S$ via adversarial training; details can be referred to in \cite{li2021towards}. 
        \item \textbf{Fair-MMD}. Fair-MMD minimizes the maximum mean discrepancy (MMD) of user latent variables given sensitive features $S$ in Multi-VAE \cite{louizos2016variational}. Specifically, we randomly select one dimension of $S$ and binarize it for the minimization.
    \end{itemize}
    \item \textbf{PS-Fairness.} Finally, we consider the following naive PS-Fair strategy, i.e., \textbf{PSF-NN}, where for each user, we calculate the similarities with all users with available $R_{b}$ measured by sensitive features. Then we select the $N$ closest neighbors, get the top $K$ unfair items, and remove them if they appear in the list. 
\end{itemize}
Finally, since the fair and unfair influences of sensitive features on observed ratings are entangled, a simple strategy to improve the fairness over the Naive model is through underfitting. Therefore, we design an early-stop baseline (which we name \textbf{CondVAE-ES}), which has the closet (rounded downwardly) N@100 on the validation users with PSF-VAE, to demonstrate that the fairness improvement of PSF-VAE is not due to simple underfitting. 

\subsubsection{\textbf{Comparison Results}} The comparison between PSF-VAE and various baselines is shown in Table \ref{tab:com_results}. The best results (compared across four classes) are shown in \textbf{bold}, and the runner-ups are \underline{underlined}. In summary, we have the following observations: \textbf{(1)} By utilizing all information in sensitive features for recommendations, CondVAE has the best recommendation performance and the worst fairness. \textbf{(2)} By simply ignoring the sensitive features, the Unawareness model (Multi-VAE) has improved fairness over the Naive model, while the recommendation performance is decreased simultaneously.  \textbf{(3)} RSs with Total Fairness further improve the fairness over Multi-VAE, since the correlations between sensitive features and observed ratings are removed from user latent variables. However, since the fair influences of sensitive features are indiscriminately discarded, they also have the worst recommendation performance. \textbf{(4)} Although PSF-NN achieves better fairness than CondVAE, the improvement is not significant. The reason could be that the nearest-neighbor strategy is too crude to model the complicated unfair influences of sensitive features on observed ratings. \textbf{(5)} PSF-VAE has much better recommendation performance than the Total Fairness models and better fairness than the Naive and Unawareness models, because PSF-VAE only blocks the unfair influence of sensitive features on ratings, while their fair effects on user interests are maximally preserved for recommendations. 

\begin{figure}
\centering
\begin{subfigure}[b]{0.41\textwidth}
\includegraphics[width=\textwidth]{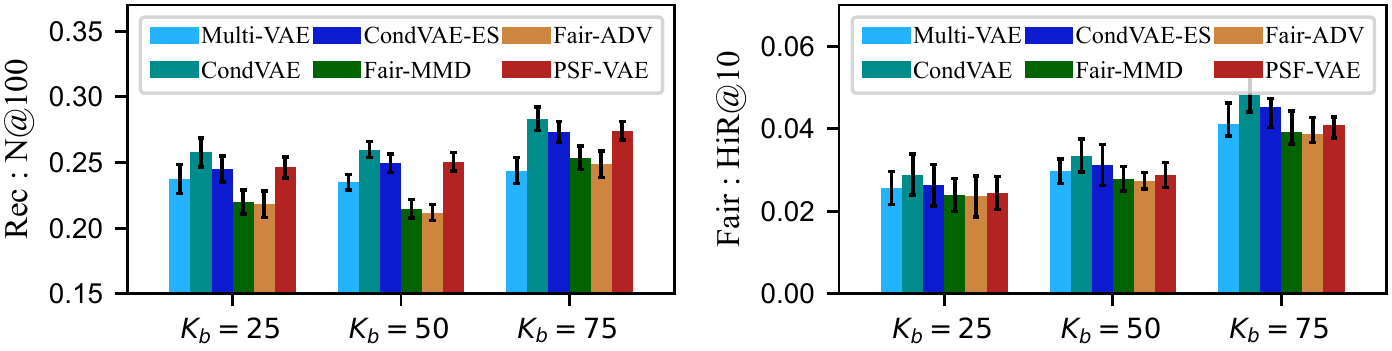}
\vspace{-5mm}
\caption{AM-VG dataset}
\vspace{2mm}
\end{subfigure}

\centering
\begin{subfigure}[b]{0.41\textwidth}
\includegraphics[width=\textwidth]{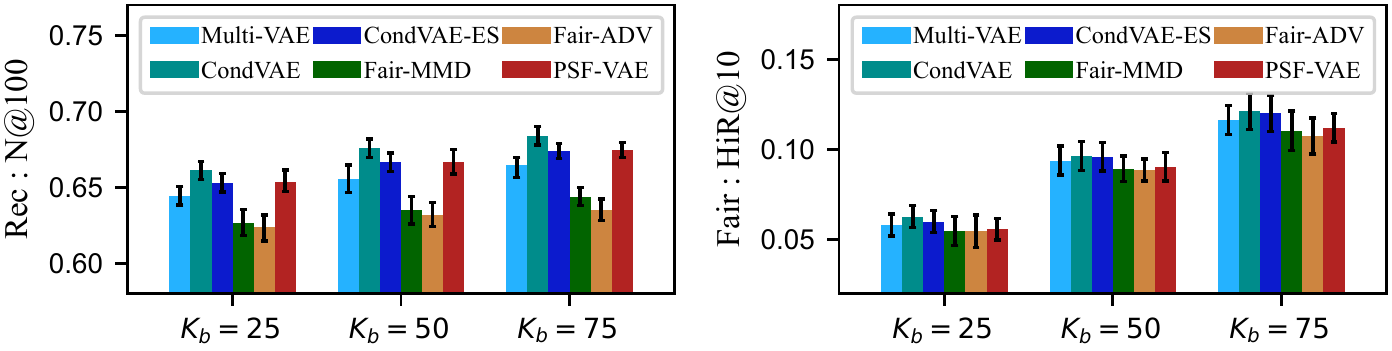}
\vspace{-5mm}
\caption{ML-1M dataset}
\end{subfigure}
\vspace{-3mm}
\caption{Comparison between PSF-VAE and baselines with different dimension of $U_{b}$$, i.e., K_{b}$, for the simulated datasets.}
\vspace{-2mm}
\label{fig:emb}
\end{figure}

In addition, we set the simulated dimension of $U_{b}$, i.e., $K_{b}$, to different values in the AM-VG and ML-1M datasets to change the relative strengths of fair and unfair causal influences of sensitive features on the observed ratings and repeat the experiments in Fig. \ref{fig:emb}. Fig. \ref{fig:emb} further demonstrates that PSF-VAE achieves a better balance between the recommendation performance and fairness compared to the Naive, Unawareness, and Total Fairness baselines.

\subsection{Ablation Study}
In this section, we compare the proposed PSF-VAE with the following variants as the ablation study to further verify its effectiveness.
\begin{itemize}[leftmargin=0.5cm]
    \item \textbf{PSF-VAE-nLat} removes the user bias variable $U_{b}$ and directly constrains the user latent variables $U$ in Multi-VAE to be independent of the observed unfair items $R_{b}$ via adversarial training. 
    \item \textbf{PSF-VAE-nWSL} removes the weakly-supervised learning module of PSF-VAE, i.e., when fitting on the biased observed ratings $R$ as Eq. (\ref{eq:elbo}), we only introduce the user bias latent variable $U_{b}$ for the subset of users with observed unfair items $R_{b}$.
    \item \textbf{PSF-VAE-nADV} removes the adversarial training module in PSF-VAE that ensures the conditional independence between latent mediators $U_{f}$ and $U_{b}$ given user sensitive features $S$. 
    \item \textbf{PSF-VAE-Mask} trains the same generative and inference networks as PSF-VAE. However, instead of learning a new model $p_{psf}(R|U_{f})$, it masks out the weights in $p_{\theta}(R|U_{f}, U_{b})$ that correspond to $U_{b}$, which leads to a new distribution $p_{masked(\theta)}(R|U_{f})$, and uses $p_{masked(\theta)}$ to make the recommendations.
\end{itemize}

\begin{table}[]
\setlength{\tabcolsep}{2pt}
\centering
\caption{Comparisons between different variants of PSF-VAE.}
\vspace{-2mm}
\label{tab:ablation_results}
\small
\begin{tabular}{lccc}
\toprule
{\textbf{AM-VG}} & \textbf{Rec: R@20} $\uparrow$ & \textbf{Rec: N@100} $\uparrow$ & \textbf{Fair: HiR@10} $\downarrow$   \\ \midrule
PSF-VAE-nLat & 0.2276 {\scriptsize $\pm$ 0.0080} & 0.2102 {\scriptsize $\pm$ 0.0045} & 0.0270 {\scriptsize $\pm$ 0.0022} \\ 
PSF-VAE-nWSL & 0.2729 {\scriptsize $\pm$ 0.0084} & 0.2543 {\scriptsize $\pm$ 0.0046} & 0.0299 {\scriptsize $\pm$ 0.0021} \\ 
PSF-VAE-nADV & 0.2721 {\scriptsize $\pm$ 0.0093} & 0.2528 {\scriptsize $\pm$ 0.0061} & 0.0297 {\scriptsize $\pm$ 0.0029} \\ 
PSF-VAE-Mask & 0.2624 {\scriptsize $\pm$ 0.0096} & 0.2463 {\scriptsize $\pm$ 0.0074} & 0.0291 {\scriptsize $\pm$ 0.0031} \\
\midrule
PSF-VAE & 0.2691 {\scriptsize $\pm$ 0.0104} & 0.2507 {\scriptsize $\pm$ 0.0075} & 0.0288 {\scriptsize $\pm$ 0.0032} \\ \bottomrule \\ 
\specialrule{0em}{-2pt}{-2pt}
\toprule{\textbf{ML-1M}} & \textbf{Rec: R@20} $\uparrow$ & \textbf{Rec: N@100} $\uparrow$ & \textbf{Fair: HiR@10} $\downarrow$ \\ \midrule
PSF-VAE-nLat & 0.5163 {\scriptsize $\pm$ 0.0152} & 0.6246 {\scriptsize $\pm$ 0.0073} & 0.0869 {\scriptsize $\pm$ 0.0083} \\ 
PSF-VAE-nWSL & 0.5647 {\scriptsize $\pm$ 0.0135} & 0.6691 {\scriptsize $\pm$ 0.0069} & 0.0932 {\scriptsize $\pm$ 0.0081} \\ 
PSF-VAE-nADV & 0.5630 {\scriptsize $\pm$ 0.0149} & 0.6687 {\scriptsize $\pm$ 0.0075} & 0.0925 {\scriptsize $\pm$ 0.0072} \\ 
PSF-VAE-Mask  & 0.5577 {\scriptsize $\pm$ 0.0132}  & 0.6659 {\scriptsize $\pm$ 0.0063}  & 0.0911 {\scriptsize $\pm$ 0.0068}\\
\midrule
PSF-VAE & 0.5601 {\scriptsize $\pm$ 0.0148} & 0.6668 {\scriptsize $\pm$ 0.0070} & 0.0904 {\scriptsize $\pm$ 0.0084} \\ \bottomrule\\ 
\specialrule{0em}{-2pt}{-2pt}
\toprule
{\textbf{LinkedIn}} & \textbf{Rec: R@20} $\uparrow$ & \textbf{Rec: N@100} $\uparrow$ & \textbf{Fair: HiR@10}  $\downarrow$   \\ \midrule
PSF-VAE-nLat & 0.1868 {\scriptsize $\pm$ 0.0048}  & 0.2832 {\scriptsize $\pm$ 0.0035} & 0.0614 {\scriptsize $\pm$ 0.0033} \\ 
PSF-VAE-nWSL & 0.2047 {\scriptsize $\pm$ 0.0041}  & 0.3009  {\scriptsize $\pm$ 0.0032} &  0.0675 {\scriptsize $\pm$ 0.0035}\\ 
PSF-VAE-nADV & 0.2032 {\scriptsize $\pm$ 0.0046}  & 0.3004 {\scriptsize $\pm$ 0.0040}  & 0.0660 {\scriptsize $\pm$ 0.0039} \\ 
PSF-VAE-Mask & 0.2016 {\scriptsize $\pm$ 0.0039} & 0.2969 {\scriptsize $\pm$ 0.0051} & 0.0654 {\scriptsize $\pm$ 0.0044} \\
\midrule
PSF-VAE   & 0.2024 {\scriptsize $\pm$ 0.0045} & 0.2987 {\scriptsize$\pm$ 0.0034} & 0.0647 {\scriptsize$\pm$ 0.0029} \\ \bottomrule  \\ 
\end{tabular}
\vspace{-10mm}
\end{table}

From Table \ref{tab:ablation_results} we can find that, PSF-VAE-nLat has the worst recommendation performance among all the variants, which shows that directly conducting adversarial training on the observed unfair items $\mathbf{r}_{b}$ is not stable, as $\mathbf{r}_{b}$ are high dimensional sparse vectors. In addition, PSF-VAE-nWSL, PSF-VAE-nADV, and PSF-VAE-Mask have worse fairness compared with PSF-VAE, with comparable recommendation performance. The results further validate the effectiveness of the weakly supervised learning and adversarial training modules of PSF-VAE to promote PS-Fairness in recommendations.

\subsection{Sensitivity Analysis}

\begin{figure}
\centering
\begin{subfigure}[b]{0.42\textwidth}
\includegraphics[width=\textwidth]{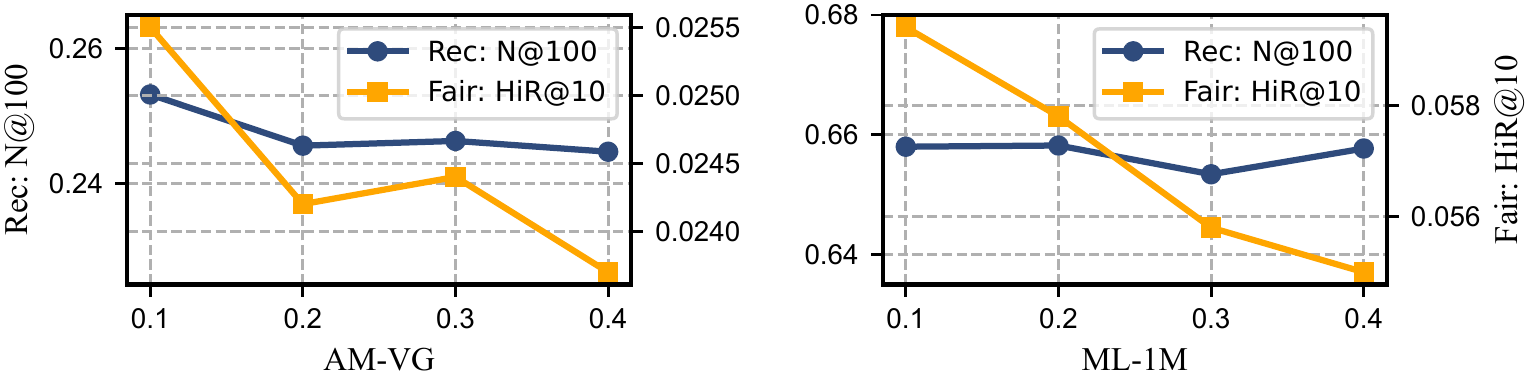}
\vspace{-5mm}
\caption{Setting: $K = 200$, $K_{b} = 25$}
\vspace{2mm}
\end{subfigure}
\begin{subfigure}[b]{0.42\textwidth}
\includegraphics[width=\textwidth]{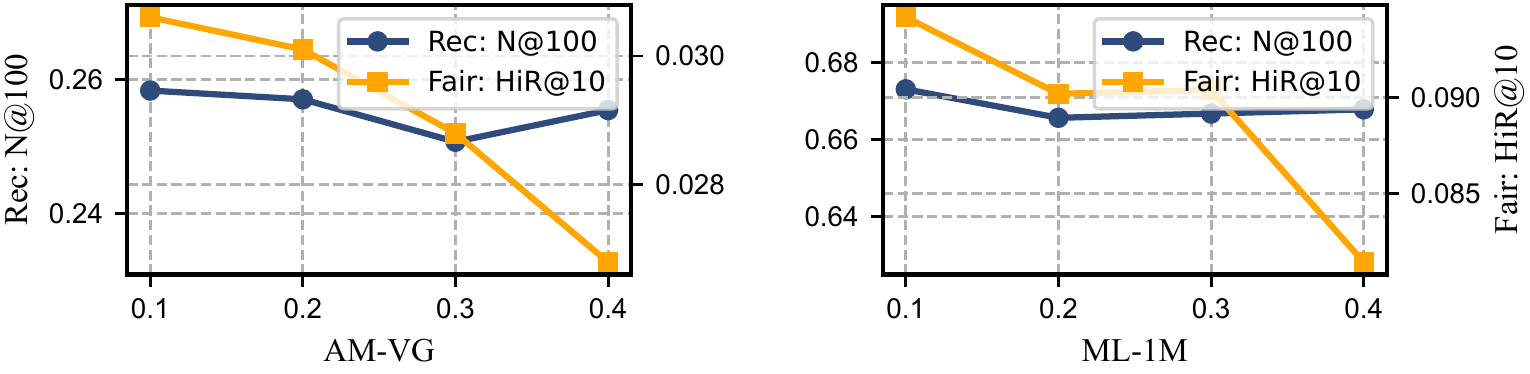}
\vspace{-5mm}
\caption{Setting: $K = 200$, $K_{b} = 50$}
\vspace{0mm}
\end{subfigure}
\caption{Sensitivity of PSF-VAE with different percentages of users with observed unfair items in AM-VG, ML-1M datasets.}
\vspace{-5mm}
\label{fig:rr}
\end{figure}

To answer \textbf{RQ 3}, we vary the mask rate of users with known unfair items in the simulated datasets, i.e., $1-c_{r}$, and plot the relations with recommendation performance and fairness in Fig. \ref{fig:rr}. From Fig. \ref{fig:rr} we can find that, the fairness of PSF-VAE generally improves with the increase of $c_{r}$, with slight negative influences on recommendation performance. This indicates that although PSF-VAE can perform well with small $c_{r}$, encouraging more users to provide feedback on unfair items can further promote PS-Fairness in recommendations.

\section{Related Work}

\noindent \textbf{Fair RSs.} Generally, fairness in RSs is a multi-stake problem, where users, items, and providers have different fairness demands. PSF-RS focuses on user-oriented fairness, which aims to fairly treat users from different demographic backgrounds \cite{zhu2018fairness,zhu2021fairness,li2021user,wei2022comprehensive}. Traditional fair RSs mainly rely on statistic parity to ensure user-oriented fairness, with metrics such as demographical parity, equalized odds, etc. \cite{calders2009building,hardt2016equality}. However, recent research indicates that the statistical discrepancy between the outcomes of different user groups may be well explained by some important non-sensitive factors \cite{khademi2019fairness,zhang2016causal,zhang2018fairness}, and algorithms that indiscriminately enforce statistical parity may still be biased against certain user groups or individuals \cite{kusner2017counterfactual,ma2022learning}.
\vspace{0.15cm}

\noindent \textbf{Causal RSs.} Causal RS is an emerging research area that reasons with the causal mechanisms underlying the observed user behaviors \cite{ma2021multi,zhu2022deep,xu2023causal}. Through a causal lens, user-oriented unfairness can be viewed as a non-confounder-induced bias due to the undesirable causal effects of sensitive features on the observed user ratings \cite{chen2020bias,zhu2023causal}. Existing causality-aware fair RSs treat all causal effects of sensitive features on ratings as unfair and remove them indiscriminately through adversarial training \cite{li2021towards} or MMD minimization \cite{louizos2016variational}. In contrast, PSF-RS preserves the fair influence of sensitive features on recommendations by identifying the fair and unfair latent mediators of sensitive features, where fairness can be achieved with the diversity of recommendations maximally preserved.

\section{Conclusions}

In this paper, we propose a path-specific fair recommender system (PSF-RS) to address the unfairness in recommendations while maximally preserving the fair influences of sensitive features on user interest. Specifically, PSF-RS summarizes all fair and unfair correlations between sensitive features and the observed ratings into two latent proxy mediators, which can be disentangled with weakly supervised variational inference based on the extremely sparse observed unfair items. To address the bias, we minimally alter the biased factual world into a hypothetically fair world, where a fair RS is learned accordingly by solving a constrained optimization problem. Extensive experiments show the effectiveness of PSF-RS.

\section*{Acknowledgments}

This work is supported by the National Science Foundation (NSF) under grants (IIS-2006844, IIS-2144209, IIS-2223769, CNS-2154962, and BCS-2228534), the Commonwealth Cyber Initiative Awards (VV-1Q23-007 and HV-2Q23-003), the JP Morgan Chase Faculty Research Award, the Cisco Faculty Research Award, the Jefferson Lab Subcontract 23-D0163, the UVA 3Cavaliers Seed Grant, and the 4-VA Collaborative Research Grant.

\balance
\bibliographystyle{ACM-Reference-Format}
\bibliography{ref}

\newpage
\appendix

\section{Definition of Causal Concepts}
\label{sec:app_causal}

\hangafter 1
\hangindent 1.5em
\noindent \textbf{Causal Graph.} A causal graph $G = (\mathcal{V}, \mathcal{E})$ is a directed acyclic graph that describes the causal relationships among the variables of interests, where $\mathcal{V}$ is the set of nodes (which represent random variables in this paper), and $\mathcal{E}$ is the set of edges, respectively. Specifically, a directed edge from variable $X$ to variable $Y$ indicates that $X$ has a causal influence on $Y$.

\hangafter 1
\hangindent 1.5em
\noindent \textbf{Structural Equations.} Each causal graph $G = (\mathcal{V}, \mathcal{E})$ can be associated with a set of structural equations $\mathcal{F}=\{p(X|Pa(X)) \mid X \in \mathcal{V}\}$, where $p(X|Pa(X))$ quantifies the causal influence of the parents nodes of $X$, i.e., $Pa(X)$, on $X$.

\hangafter 1
\hangindent 1.5em
\noindent \textbf{Causal Path.} A causal path $P$ between variables $X$ and $Y$ is a sequence of edges (from $X$ to $Y$) in $\mathcal{E}$ such that each edge starts with the node that ends the previous edge. A directed causal path is a causal path whose edges point in the same direction.

\hangafter 1
\hangindent 1.5em
\noindent \textbf{Mediator/Mediate.} In a directed causal path  $P$ between $X$ and $Y$, e.g., $X \rightarrow M \rightarrow Y$, any intermediate node $M$ is a mediator, where the causal effects of $X$ on $Y$ are mediated by $M$.

\hangafter 1
\hangindent 1.5em
\noindent \textbf{Block/Unblock.}
If conditioning on $M=m$ blocks the causal path $P$ between $X$ and $Y$, no dependence (both causal and non-causal ones) can be passed from $X$ to $Y$ along the path $P$ when $M$ is known (see \cite{glymour2016causal} for a formal definition). Otherwise, we say that conditioning on $M=m$ unblocks the causal path $P$.

\hangafter 1
\hangindent 1.5em
\noindent \textbf{Intervention.}
Given a causal graph $G$, we can conduct interventions on a variable $X$, which means that we set $X$ to a value $x$ regardless of its observed values as well as the values of its parents $Pa(X)$. If unspecified, the intervention is conducted upon the whole population, but we can also conduct the intervention conditional on $C=c$, which means that we set $X = x$ on the sub-population specified by the conditions.

\hangafter 1
\hangindent 1.5em
\noindent \textbf{Potential Outcome.} Potential outcomes can be used to formalize the definition of interventions. Specifically, we define the potential outcome $Y_{X \leftarrow x}(i)$ as the value of $Y$ for unit $i$ had $X$ been $x$. Based on $Y_{X \leftarrow x}(i)$, we can further define the potential outcome \textbf{random variable} $Y_{X \leftarrow x}$ to denote the unconditional intervention that set $X=x$ uniformly upon the population. Furthermore, the conditional potential outcome random variable $Y_{X \leftarrow x} | C=c$ can be used to denote the intervention conducted upon the sub-population specified by the condition $C=c$. 

\hangafter 1
\hangindent 1.5em
\noindent  \textbf{Counterfactuals}. For $Y_{X \leftarrow x} | C=c$, when $C = X$ and $c = x^{\prime}$, the conditional potential random variable $Y_{X \leftarrow x} | X=x^{\prime}$ can be used to define the counterfactual distribution of $Y$ had $X$ for the units with the factual value of $X=x^{\prime}$ been set to a counterfactual value $x$. The above analysis also applies to the \textbf{Nested Potential Outcome} introduced in Definition \ref{def:npo}.


\section{Theoretical Analysis}
\subsection{Proof of Identification of PS-Bias in Eq. (\ref{eq:psb})}
\label{sec:proof_identify}
\begin{assumption}\textbf{Sequential Ignorability \cite{imai2010general}.}  \\
\noindent \textbf{Step 1.} We assume that given $X$, the sensitive features $S$ are ignorable for the mediators $U_{f}$, $U_{b}$ and user ratings $R$ as follows: 
\begin{equation}
\label{eq:step1}
U_{f, S \leftarrow \mathbf{s}}, U_{b, S \leftarrow \mathbf{s}^{\prime}}, \ R_{S \leftarrow \mathbf{s}^{\prime},  U_{f} \leftarrow \mathbf{u}_{f}, U_{b} \leftarrow \mathbf{u}^{\prime}_{b}  } \Perp S | X.
\end{equation}
\textbf{Step 2.} We also assume that given $X$, the post-interventional mediators $U_{f, S \leftarrow \mathbf{s}}$, $U_{b, S \leftarrow \mathbf{s}^{\prime}}$ are ignorable for the user ratings $R$ as follows:
\begin{equation}
\label{eq:step2}
R_{S \leftarrow \mathbf{s}^{\prime}, U_{f} \leftarrow \mathbf{u}_{f}, U_{b} \leftarrow \mathbf{u}^{\prime}_{b}} \Perp U_{f, S \leftarrow \mathbf{s}}, U_{b, S \leftarrow \mathbf{s}^{\prime}} | X.    
\end{equation}
\label{ass:seq_ign}
\end{assumption}

\noindent The difference between the potential outcome $R_{S \leftarrow \mathbf{s}^{\prime}, U_{f} \leftarrow \mathbf{u}_{f}, U_{b} \leftarrow \mathbf{u}^{\prime}_{b}}$ and the nested potential outcome $R_{S \leftarrow \mathbf{s}^{\prime}}(U_{f, S \leftarrow \mathbf{s}},  U_{b, S \leftarrow \mathbf{s}^{\prime}})$ lies in the fact that the former directly sets the mediators $U_{f}$ and $U_{b}$ to the values $\mathbf{u}_{f}$ and $\mathbf{u}^{\prime}_{b}$, whereas the latter conducts interventions on $S$ by setting $S$ to $\mathbf{s}$ and $\mathbf{s}^{\prime}$ and let them influence $U_{f}$ and $U_{b}$. 

The sequential ignorability assumption holds for the causal graph specified in Fig. \ref{fig:causal_graph}, because there are no unobserved confounders for the causal paths $S\rightarrow U_{f}$, $S \rightarrow U_{b}$ and $S \rightarrow R$ (and thus Eq. (\ref{eq:step1}) holds) and $U_{f} \rightarrow R$ and $U_{b} \rightarrow R$ (and thus Eq. (\ref{eq:step2}) holds).

\subsubsection{\textbf{Proof}} Based on the sequential ignorability assumption defined above, Eq. (\ref{eq:bias_nested_dist}) can be proved with six steps as follows:\begin{equation}
\begin{aligned}
&\mathbb{E}\left[R_{S \leftarrow \mathbf{s}^{\prime}}\left(U_{f, S \leftarrow \mathbf{s}},  U_{b, S \leftarrow \mathbf{s}^{\prime}} \right) = \mathbf{r}  \big | X=\mathbf{x}, S=\mathbf{s}\right]\\
\overset{(a)}{\longeq} \int_{\mathbf{r}, \mathbf{u}_{f}  , \mathbf{u}^{\prime}_{b}} & p\Big(R_{S \leftarrow \mathbf{s}^{\prime}}\left(U_{f, S \leftarrow \mathbf{s}},  U_{b, S \leftarrow \mathbf{s}^{\prime}}\right) = \mathbf{r} \Big| X=\mathbf{x}, S=\mathbf{s}, U_{f, S \leftarrow \mathbf{s}} = \mathbf{u}_{f}  , \\
& \quad U_{b, S \leftarrow \mathbf{s}^{\prime}} = \mathbf{u}^{\prime}_b \Big) \cdot p\left(U_{f, S \leftarrow \mathbf{s}} = \mathbf{u}_{f}   \big | X=\mathbf{x}, S=\mathbf{s}\right) \cdot \\
&p\left(U_{b, S \leftarrow \mathbf{s}^{\prime}} = \mathbf{u}^{\prime}_b \big | X=\mathbf{x}, S=\mathbf{s} \right)  \cdot \mathbf{r}\\
\overset{(b)}{\longeq} \int_{\mathbf{r}, \mathbf{u}_{f}  , \mathbf{u}^{\prime}_{b}}  & p\Big(R_{S \leftarrow \mathbf{s}^{\prime}, U_{f} \leftarrow \mathbf{u}_{f}  , U_{b} \leftarrow \mathbf{u}^{\prime}_{b}} = \mathbf{r} \Big | X=\mathbf{x}, S=\mathbf{s}, U_{f, S \leftarrow \mathbf{s}} = \mathbf{u}_{f}  , \\
&\quad U_{b, S \leftarrow \mathbf{s}^{\prime}} = \mathbf{u}^{\prime}_{b}\Big) \cdot p\left(U_{f, S \leftarrow \mathbf{s}} = \mathbf{u}_{f}   \big | X=\mathbf{x}, S=\mathbf{s} \right) \cdot \\
&p\left(U_{b, S \leftarrow \mathbf{s}^{\prime}} = \mathbf{u}^{\prime}_b \big | X=\mathbf{x}, S=\mathbf{s} \right) \cdot \mathbf{r} \\
\overset{(c)}{\longeq} \int_{\mathbf{r}, \mathbf{u}_{f}  , \mathbf{u}^{\prime}_{b}} & p\left(R_{S \leftarrow \mathbf{s}^{\prime}, U_{f} \leftarrow \mathbf{u}_{f}  , U_{b} \leftarrow \mathbf{u}^{\prime}_{b}} =\mathbf{r} \big | X=\mathbf{x}, S=\mathbf{s} \right) \cdot \mathbf{r} \cdot \\
&p\left(U_{f, S \leftarrow \mathbf{s}} = \mathbf{u}_{f}   \big | X=\mathbf{x}, S=\mathbf{s}\right) \cdot p\left(U_{b, S \leftarrow \mathbf{s}^{\prime}} = \mathbf{u}^{\prime}_b \big | X=\mathbf{x}, S=\mathbf{s}\right)\\
\overset{(d)}{\longeq} \int_{\mathbf{r}, \mathbf{u}_{f}  , \mathbf{u}^{\prime}_{b}} & p\left(R_{S \leftarrow \mathbf{s}^{\prime}, U_{f} \leftarrow \mathbf{u}_{f}  , U_{b} \leftarrow \mathbf{u}^{\prime}_{b}} =\mathbf{r} \big | X=\mathbf{x}\right) \cdot\\
& p\left(U_{f, S \leftarrow \mathbf{s}} = \mathbf{u}_{f}   \big | X=\mathbf{x}\right) \cdot p\left(U_{b, S \leftarrow \mathbf{s}^{\prime}} = \mathbf{u}^{\prime}_b \big | X=\mathbf{x}\right) \cdot \mathbf{r}\\
\overset{(e)}{\longeq} \int_{\mathbf{r}, \mathbf{u}_{f}  , \mathbf{u}^{\prime}_{b}} & p(\mathbf{r} | \mathbf{u}_f, \mathbf{u}^{\prime}_b, \mathbf{s}^{\prime}, \mathbf{x}) \cdot p(\mathbf{u}_f | \mathbf{s}, \mathbf{x}) \cdot p(\mathbf{u}^{\prime}_b | \mathbf{s}^{\prime}, \mathbf{x}) \cdot p(\mathbf{x}) \cdot \mathbf{r}\\
\overset{(f)}{\longeq} \int_{\mathbf{r}, \mathbf{u}_{f}  , \mathbf{u}^{\prime}_{b}} & p(\mathbf{r} | \mathbf{u}_f, \mathbf{u}^{\prime}_b(\mathbf{s}^{\prime})) \cdot p(\mathbf{u}_f | \mathbf{s}, \mathbf{x}) \cdot p(\mathbf{u}^{\prime}_b | \mathbf{s}^{\prime}) \cdot p(\mathbf{x}) \cdot \mathbf{r}.
\end{aligned}
\end{equation}
Step (a) is based on the total probability theory; step (b) is based on the consistency rule of counterfactuals \cite{rubin1980randomization}; step (c) is based on the second step of sequential ignorability; steps (d)(e) are based on the first step of sequential ignorability; and step (f) is based on the conditional independence assumptions implied by the causal graph in Fig. \ref{fig:causal_graph}. Similar procedures can be used to prove the identification of Eq. (\ref{eq:nested_dist}), where Eq. (\ref{eq:psb}) can be calculated as Eq. (\ref{eq:bias_nested_dist}) - Eq. (\ref{eq:nested_dist}).

\subsection{\textbf{PS-Bias for RS Models with Constraints}}
\label{sec:ps-bias-generalize}
In section \ref{sec:naive_rs}, we have introduced the PS-Bias of the naive RSs that predict new ratings according to the exact causal mechanism that generates the biased observed ratings. This section generalizes the PS-Bias for RS models with extra constraints, which serves as the basis for proving the PS-Bias for existing fair RSs and PSF-RS.

We note that the causal mechanism that
generates the observed ratings is composed of three structural equations: $\mathcal{F} = \{p(R|U_{f}, U_{b})$, $p(U_{f}|S, X)$, $p(U_{b}|S)\}$, which induces the causal graph in Fig. \ref{fig:causal_graph} by setting the variables on the RHS of $p \in \mathcal{F}$ as the parents and the variable on the LHS as the child. An RS model with extra constraints \textbf{can be viewed as} generating ratings in two steps: \textbf{(1)} Certain structural equations $p$ in $\mathcal{F}$ are \textit{minimally} changed to $p_{model}$ according to the constraints (where the irrelevant ones remain intact). We use $\mathcal{F}_{model}$ to denote the new set of structural equations, which induces a new causal graph (e.g., Figs. \ref{fig:naive-rs}-(b) and \ref{fig:fair-rs}-(b)). \textbf{(2)} Ratings are generated according to the newly-induced causal model. Therefore, PS-Bias for an RS with constraints can be calculated as the path-specific effects of sensitive features $S$ on ratings $R$ along the unfair paths of the \textbf{newly-induced causal model}. 

\subsection{Proof of Zero PS-Bias for Existing Fair RSs}
\label{sec:nb_efrs}
\subsubsection{\textbf{Further Analysis}}
Existing fair RSs constrain the user latent variables $U$ to be independent of the user sensitive features $S$ as Eq. (\ref{eq:efairness_const}). To satisfy such a constraint, we need to change at least two structural equations in $\mathcal{F}$, i.e., $p(U_{f}|S,X)$, $p(U_{b}|S)$ into $p_{ef}(U_{f}|X)$, $p_{ef}(U_{b})$ (although in practice, when maximizing the likelihood of observed ratings, $p(R|U_{f}, U_{b})$ will also be changed into $p_{ef}(R|U_{f}, U_{b})$ since the distributions of $U_{f}, U_{b}$ are altered), where the causal structure $p(U_{f}|S,X)$ necessary for recommendation diversity is inevitably lost. We use $PSBias^{**}(\mathbf{x},\mathbf{s},\mathbf{s}^{\prime})$ to denote the PS-Bias of the altered causal model induced by existing fair RSs.

\subsubsection{\textbf{Proof.}} $PSBias^{**}(\mathbf{x},\mathbf{s},\mathbf{s}^{\prime})$ can be calculated by substituting the three $p_{ef}$ terms introduced above for the $p$ terms in Eq. (\ref{eq:psb}). After the substitution, the first expectation term becomes
\begin{equation}
\label{eq:after11}
\begin{aligned}
 \mathbb{E}^{**}&\left[R_{S \leftarrow \mathbf{s}^{\prime}}\left(U_{f, S \leftarrow \mathbf{s}},  U_{b, S \leftarrow \mathbf{s}^{\prime}}\right) \Big | X=\mathbf{x}, S=\mathbf{s} \right] \\ 
 &=\int _ {\mathbf{r}, \mathbf{u}_f, \mathbf{u}^{\prime}_b} p_{ef}(\mathbf{r} | \mathbf{u}_f, \mathbf{u}^{\prime}_b) \cdot p_{ef}(\mathbf{u}_f | \mathbf{x}) \cdot p_{ef}(\mathbf{u}^{\prime}_b) \cdot \mathbf{r}. \\
\end{aligned}
\end{equation}
Similarly, the second expectation term becomes
\begin{equation}
\label{eq:after12}
\begin{aligned}
 \mathbb{E}^{**}&\left[R_{S \leftarrow \mathbf{s}}\left(U_{f, S \leftarrow \mathbf{s}},  U_{b, S \leftarrow \mathbf{s}}\right)=\mathbf{r} \Big | X=\mathbf{x}, S=\mathbf{s} \right] \\ 
 &=\int _ {\mathbf{r}, \mathbf{u}_f, \mathbf{u}_b} p_{ef}(\mathbf{r} | \mathbf{u}_f, \mathbf{u}_b) \cdot p_{ef}(\mathbf{u}_f | \mathbf{x}) \cdot p_{ef}( \mathbf{u}_b) \cdot \mathbf{r}. \\
\end{aligned}
\end{equation}
Since $PSBias^{**}(\mathbf{x},\mathbf{s},\mathbf{s}^{\prime})$ = Eq. (\ref{eq:after11}) - Eq. (\ref{eq:after12}), the equality of Eqs. (\ref{eq:after11}) and (\ref{eq:after12}) proves that $PSBias^{**}(\mathbf{x},\mathbf{s},\mathbf{s}^{\prime}) = 0$ for existing fair RSs.

\subsection{Proof of Zero PS-Bias for PSF-RS}
\label{sec:proof_zero_bias}

In the hypothetically fair world induced by the proposed PSF-RS, $p_{psf}(R|U_{f})$ is substituted for $p(R|U_{f}, U_{b})$ in $\mathcal{F}$ while other causal mechanisms invariant to the RS remain unchanged. Similarly, the first expectation term in $PSBias^{*}(\mathbf{x},\mathbf{s},\mathbf{s}^{\prime})$ can be calculated as

\begin{equation}
\label{eq:after1}
\begin{aligned}
 \mathbb{E}^{*}&\left[R_{S \leftarrow \mathbf{s}^{\prime}}\left(U_{f, S \leftarrow \mathbf{s}},  U_{b, S \leftarrow \mathbf{s}^{\prime}}\right) \Big | X=\mathbf{x}, S=\mathbf{s} \right] \\ 
 &=\int _ {\mathbf{r}, \mathbf{u}_f, \mathbf{u}^{\prime}_b} p_{psf}(\mathbf{r} | \mathbf{u}_f) \cdot p(\mathbf{u}_f | \mathbf{s}, \mathbf{x}) \cdot p(\mathbf{u}^{\prime}_b | \mathbf{s}^{\prime}) \cdot \mathbf{r} \\
 &= \int _ {\mathbf{u}^{\prime}_b} p(\mathbf{u}^{\prime}_b | \mathbf{s}^{\prime}) \int _ {\mathbf{r}, \mathbf{u}_f} p_{psf}(\mathbf{r} | \mathbf{u}_f) \cdot p(\mathbf{u}_f | \mathbf{s}, \mathbf{x}) \cdot \mathbf{r} \\
 &= \int _ {\mathbf{r}, \mathbf{u}_f} p_{psf}(\mathbf{r} | \mathbf{u}_f) \cdot p(\mathbf{u}_f | \mathbf{s}, \mathbf{x}) \cdot \mathbf{r}.
\end{aligned}
\end{equation}
Furthermore, the second expectation term becomes
\begin{equation}
\label{eq:after2}
\begin{aligned}
 \mathbb{E}^{*}&\left[R_{S \leftarrow \mathbf{s}}\left(U_{f, S \leftarrow \mathbf{s}},  U_{b, S \leftarrow \mathbf{s}}\right)=\mathbf{r} \Big | X=\mathbf{x}, S=\mathbf{s} \right] \\ 
 &=\int _ {\mathbf{r}, \mathbf{u}_f, \mathbf{u}_b} p_{psf}(\mathbf{r} | \mathbf{u}_f) \cdot p(\mathbf{u}_f | \mathbf{s}, \mathbf{x}) \cdot p(\mathbf{u}_b | \mathbf{s}) \cdot \mathbf{r} \\
 &=\int _ {\mathbf{r}, \mathbf{u}_f} p_{psf}(\mathbf{r} | \mathbf{u}_f) \cdot p(\mathbf{u}_f | \mathbf{s}, \mathbf{x}) \cdot \mathbf{r}.
\end{aligned}
\end{equation}
Since $PSBias^{*}(\mathbf{x},\mathbf{s},\mathbf{s}^{\prime})$ = Eq. (\ref{eq:after1}) - Eq. (\ref{eq:after2}), the equality of the RHS of Eqs. (\ref{eq:after1}) and (\ref{eq:after2}) proves that $PSBias^{*}(\mathbf{x},\mathbf{s},\mathbf{s}^{\prime}) = 0$ for PSF-RS.

\subsection{Proof of ELBO for PSF-VAE}
\label{sec:elbo_proof}
In this section, we prove the ELBO of PSF-VAE in Eq. (\ref{eq:elbo}) as follows:
\begin{equation}
\label{eq:elbo_proof}
\begin{aligned}
\ln p_{\theta}&(R,  \ R_{b}|S,X) = \ln \int_{U_{f}, U_{b}} p_{\theta}(R, R_{b},U_{f}, U_{b}|S,X) \\
    = &\ \ln \int_{U_{f}, U_{b}} q_{\phi}(U_{f}, U_{b}|R, S, X) \cdot \frac{p_{\theta}(R, R_{b},U_{f}, U_{b}|S,X)}{q_{\phi}(U_{f}, U_{b}|R, S, X)} \\
    \overset{(a)}{\ge} &\ \int_{U_{f}, U_{b}} q_{\phi}(U_{f}, U_{b}|R, S, X) \cdot \ln \frac{p_{\theta}(R, R_{b},U_{f}, U_{b}|S,X)}{q_{\phi}(U_{f}, U_{b}|R, S, X)} \\
     =&\ \int_{U_{f}, U_{b}} q_{\phi}(U_{f}, U_{b}|R, S, X) \cdot \ln \frac{p_{\theta}(U_{f}, U_{b}|S,X)\cdot p_{\theta}(R, R_{b} | U_{f}, U_{b})}{q_{\phi}(U_{f}, U_{b}|R, S, X)} \\
     =&\ \mathbb{E}_{q_{\phi}(U_{f}, U_{b}|R, S, X)}\left[\ln \frac{p_{\theta}(U_{f}, U_{b}|S,X)\cdot p_{\theta}(R, R_{b} | U_{f}, U_{b})}{q_{\phi}(U_{f}, U_{b}|R, S, X)}\right] \\
     =&\ \mathbb{E}_{q_{\phi}(U_{f}, U_{b}|R, S, X)}\left[\ln  p_{\theta}(R, R_{b} | U_{f}, U_{b}) \right] + \\  &\ \mathbb{E}_{q_{\phi}(U_{f}, U_{b}|R, S, X)}\left[\frac{p_{\theta}(U_{f}, U_{b}|S,X)}{q_{\phi}(U_{f}, U_{b}|R, S, X)}\right]\\
     =&\ \mathbb{E}_{q_{\phi}(U_{f}, U_{b} | R, S, X)}[\ln p_{\theta}(R | U_{f}, U_{b})] + \mathbb{E}_{q_{\phi}(U_{b} | R, S)}[\ln p_{\theta}(R_{b} | U_{b})] \\
    -&\ \mathbb{KL}[q_{\phi}(U_{f} | R, S, X) \ || \ p_{\theta}(U_{f}|S,X)] - \mathbb{KL}[q_{\phi}(U_{b} | R, S) \ || \ p_{\theta}(U_{b}|S)],
\end{aligned}
\end{equation}
where step (a) is the application of Jensen's inequality, and the final step is based on the conditional independence assumptions implied by the causal graph in Fig.  \ref{fig:causal_graph}, which leads to the ELBO in Eq. (\ref{eq:elbo}). 

We can further show that the difference between the ELBO and the log evidence $\ln p_{\theta}(R, R_{b}|S,X)$ is exactly the KL-divergence between variational posterior $q_{\phi}(U_{f}, U_{b}|R,S,X) = q_{\phi}(U_{f}|R,S,X) \times q_{\phi}(U_{b}|R,S)$ and the true posterior $p_{\theta}(U_{f}, U_{b}|R,R_{b},S,X)$. To prove this, we can add the KL term to the RHS of (a) in Eq. (\ref{eq:elbo_proof}) as follows:
\begin{equation}
\label{eq:kl2elbo}
\begin{aligned}
    & (a) + \mathbb{KL}[q_{\phi}(U_{f}, U_{b} | R, S, X) \ || \ p_{\theta}(U_{f}, U_{b} | R, R_{b},S,X)]  \\
    =\ &\mathbb{E}_{q_{\phi}(U_{f}, U_{b}|R, S, X)}\left[\ln \frac{p_{\theta}(R, R_{b}, U_{f}, U_{b}|S,X)}{q_{\phi}(U_{f}, U_{b}|R, S, X)} \cdot \frac{q_{\phi}(U_{f}, U_{b} | R, S, X)}{p_{\theta}(U_{f}, U_{b} | R, R_{b},S,X)}\right] \\ 
    =\ & \mathbb{E}_{q_{\phi}(U_{f}, U_{b}|R, S, X)}\left[\ln \frac{p_{\theta}(R, R_{b}, U_{f}, U_{b} | S,X)}{p_{\theta}(U_{f}, U_{b} | R, R_{b},S,X)}\right] \\
    =\ & \mathbb{E}_{q_{\phi}(U_{f}, U_{b}|R, S, X)}\left[\ln p_{\theta}(R, R_{b}|S,X)\right] = \ln p_{\theta}(R, R_{b}|S,X),
\end{aligned}
\end{equation}
where the RHS of Eq. (\ref{eq:kl2elbo}) is the log evidence $\ln p_{\theta}(R, R_{b}|S,X)$. This further proves our claim that minimizing the KL divergence between the variational posteriors defined by PSF-VAE and the true posteriors is equivalent to maximizing the ELBO as Eq. (\ref{eq:elbo}).

\end{document}